\documentclass{jfm}
\usepackage{graphicx}
\usepackage{epstopdf, epsfig}
\usepackage{hyperref}
\usepackage[usenames,dvipsnames,svgnames,table]{xcolor}
\usepackage{mathptmx} 
\usepackage[utf8]{inputenc}
\usepackage{newtxmath}
\usepackage{pgfplots}
\usepackage{natbib}
\usepackage{microtype}
\usepackage{bm}

\hypersetup{
colorlinks=true,
linkcolor=blue,
urlcolor=blue,
filecolor=blue,
citecolor=blue
}

\makeatletter

\patchcmd{\NAT@citex}
  {\@citea\NAT@hyper@{%
     \NAT@nmfmt{\NAT@nm}%
     \hyper@natlinkbreak{\NAT@aysep\NAT@spacechar}{\@citeb\@extra@b@citeb}%
     \NAT@date}}
  {\@citea\NAT@nmfmt{\NAT@nm}%
   \NAT@aysep\NAT@spacechar\NAT@hyper@{\NAT@date}}{}{}

\patchcmd{\NAT@citex}
  {\@citea\NAT@hyper@{%
     \NAT@nmfmt{\NAT@nm}%
     \hyper@natlinkbreak{\NAT@spacechar\NAT@@open\if*#1*\else#1\NAT@spacechar\fi}%
       {\@citeb\@extra@b@citeb}%
     \NAT@date}}
  {\@citea\NAT@nmfmt{\NAT@nm}%
   \NAT@spacechar\NAT@@open\if*#1*\else#1\NAT@spacechar\fi\NAT@hyper@{\NAT@date}}
  {}{}

\makeatother

\defcitealias{companion}{GMS}

\newcommand{\vect}[1]{\boldsymbol{#1}}

\newcommand\Ri{\mbox{\textit{Ri}}} 
\newcommand\Sc{\mbox{\textit{Sc}}} 
\newcommand\im{\mathrm{i}\mkern1mu} 

\title{Exact coherent structures with dilute particle suspensions}
\shorttitle{Exact coherent structures with dilute particle suspensions}

\shortauthor{J. Langham and A.J. Hogg}
\author{Jake Langham\aff{1,2}
\corresp{\email{jacob.langham@manchester.ac.uk}},
 \and Andrew J. Hogg\aff{2}}
\affiliation{\aff{1}Department of Mathematics and Manchester Centre for Nonlinear Dynamics, University of Manchester, Oxford Road, Manchester M13 9PL, UK
    \aff{2}School of Mathematics, Fry Building, 
University of Bristol, Bristol, BS8 1UG, UK
}

\begin{document}
\maketitle

\begin{abstract}
The physics of settling suspensions under shear are investigated by theoretical
and numerical analyses of unstable equilibrium solutions to the
incompressible Navier-Stokes equations, coupled with an
advection--diffusion--settling
equation for a dilute phase of particles.  Two cases are considered: the
`passive scalar' regime, in which the sediment 
is advected by the fluid motion, but concentrations are 
too dilute to affect the
flow; and the `stratified' regime, where nonuniform vertical
distribution of sediment due to particle settling leads to a bulk
stratification that feeds back on the flow via buoyancy.  In the passive
regime, we characterise the structure of the resultant sediment concentration fields
and derive formulae for transport fluxes of sediment with asymptotically low
and high settling velocities.  In the stratified regime, parametric
continuation is employed to explore the dependence of states upon the bulk
Richardson number $\Ri_b$. Symmetry breaking in the governing equations 
leads to travelling wave solutions with a rich bifurcation structure.
The maximum $\Ri_b$ attained by these states depends non-monotonically on
settling velocity and obeys asymptotic scalings that have also been
observed to capture the dependence of the
laminar--turbulent boundary in direct numerical simulations.
\end{abstract}
\section{Introduction}
Velocity fluctuations within a turbulent fluid may hold small particles
aloft long enough to transport them long distances.
The resulting flows drive particulate transport within the oceans and
atmosphere, forming a key part of the global sediment
cycle and underpinning natural phenomena such as
underwater turbidity currents~\citep{Meiburg2010}, river
suspensions~\citep{Martinelli1989} and the
atmospheric dispersion of dust and other pollutants~\citep{Bursik1996,Schepanski2018}.
Consequently, the problem of understanding and modelling sediment transport by
turbulence impinges upon many topics of practical interest in science and
engineering~\citep{Dyer1988,Elfrink2002,Gislason2006,Dethier2022}. Since natural particles are typically denser than
their carrier fluid, they preferentially concentrate in the direction of gravity,
causing the local density of fluid--sediment parcels to increase towards the
base of the flow.
This stratifies the mixture and extracts energy from the turbulent eddies that
support the
suspension~\citep{Villaret1991,Winterwerp2001,Cantero2009a,Huang2022}. In some
cases, this effect can be
sufficient to fully or partially laminarise the flow, thereby collapsing the
suspension~\citep{Cantero2012b,Shringarpure2012}.

The study of turbulent particle transport and the related phenomenon of
stratified damping may be coarse grained by averaging over fluctuations within a
particular flow -- an approach which captures the most important global balances
in the system~\citep[see e.g.][]{Garcia1991,Fredsoe1992,Wright2004}.  However,
it is the aggregate effect of countless individual coherent motions within the
flow that ultimately maintains a sediment suspension. Consequently, there exists
a large body of research dedicated to observing, characterising and analysing
discrete flow structures in the turbulent boundary
layer~\citep{Kline1967,Corino1969,Robinson1991} and assessing their role in
mobilising and transporting sediment, which may be inferred via experimental
measurements and computer
simulations~\citep{Grass1971,Jackson1976,Sumer1978,Dyer1988,Nelson1995,Cellino2004,Shringarpure2012,Salinas2021b,Salinas2021a}.
In particular, studies have highlighted the central role that vortices play in
ejecting particles upwards, away from the underlying boundary and into the bulk
of the flow, thereby facilitating their transport downstream. Though it was
proposed many years ago by \cite{Dyer1988} that knowledge of this mechanism
could be used to improve upon the poor performance empirical sediment transport
formulae, there has been scant progress in this direction to-date.

It is within this context that we position the present contribution.
However, rather than investigating sediment-laden turbulence directly (by, for
example, measuring correlations between unsteady flow variables), we 
study a class of laminar vortex states (`exact coherent structures'), which are
invariant solutions of the Navier-Stokes equations coupled, in this case, with
an equation that describes the motion of a dilute particle
phase.
The relevance of these flow states to turbulence stems from the fact that, when
the governing equations are viewed as a dynamical system, they can lie embedded
within the turbulent part of phase space, or along surfaces that separate
laminar and turbulent conditions, and are only mildly unstable (in the sense of
possessing a low number of unstable eigenmodes relative to the high
dimensionality of the system). This means that during the course of a turbulent
flow's evolution it can make frequent, but transient, visits to these states,
which locally guide the dynamics along their stable and unstable manifolds.
This conceptual picture is clearest in small computational domains at
transitionally turbulent Reynolds numbers, where these solutions have been most
commonly studied~\citep{Gibson2008,Cvitanovic2010}.  However, it is widely
argued that unstable invariant solutions form the essential building blocks of
more complex spatially extended flows, as well as underpinning the near-wall
coherent motions that produce
and regenerate turbulence~\citep{Kawahara2001,Waleffe2001,Itano2009,Kawahara2012,Shekar2018}.

Given the mature state of the established theory for homogeneous flows, it is
reasonable to expect that these states are necessary flow structures for
supporting sediment suspensions, with properties that capture aspects of the
turbulence itself. This is useful, since individual invariant solutions, being
far simpler than a chaotic turbulent flow trajectory, are amenable to direct
mathematical analysis~\citep{Hall2010,Deguchi2014}. Furthermore, their
vortical structure provides an exact analogue to the transient coherent
structures that have long been observed to lift particles vertically and
downstream in turbulent shear flows~\citep{Jackson1976,Sumer1978}. Consequently,
understanding their role, even in simple configurations, may be a necessary
first step en route to improving sediment transport models which currently
neglect flow structures in favour of bulk statistical averages.

However, despite over three decades of research on these states in a variety of canonical
homogeneous flows, little is known about their interaction with sediment.
\cite{Pestana2020} simulated the consequences of adding finite-sized dense spherical
particles to a time-invariant solution in plane Couette flow under stable conditions,
using an immersed boundary method to capture the mutual coupling between
particle motion and the flow velocity field. They established that the essential
structure of the equilibrium state is robust to the presence of the particles,
which are guided by streamwise-aligned vortices toward regions of below average
streamwise velocity.  The case of small, yet relatively dense, inertialess
particles is yet to be addressed.  We do so herein, treating the sediment as a
continuous concentration field, as opposed to individual point particles.  This
physical description has been most widely employed to investigate the fluid
dynamics of submarine turbidity currents via direct numerical
simulations~(DNS)~\citep[e.g.][]{Necker2002,Cantero2009a}, though it retains
relevance for flows in other geophysical contexts such as river suspensions,
dust clouds and coastal swash zones, when particle concentrations are low.

The flow physics of dilute suspensions is closely related to density
stratified flows that obey the Boussinesq approximation.  These systems depend
on three dimensionless parameters: the Reynolds number $\Rey$, the bulk
Richardson number $\Ri_b$ and the Schmidt (or Prandtl) number $\Sc$, with the
latter two quantities dictating the strength of the buoyancy effect and the
ratio of the fluid viscosity to concentration (or thermal) diffusivity,
respectively.  Previous authors have explored the way in which invariant states
subject to a linear background stratification are modified as these parameters
are varied.  Among the findings are: the numerical and theoretical determination
of the regimes at which homogeneous equilibria become affected by and ultimately
disrupted by stratification~\citep{Eaves2015,Olvera2017,Deguchi2017}, the
localisation of these states away from regions of high density
gradients~\citep{Olvera2017,Langham2020} and the critical role that the Schmidt
number plays in controlling the structure of the density
field~\citep{Langham2020,Parker2021}. Invariant states have also been
linked to fundamental instabilities of these
systems~\citep{Olvera2017,Lucas2017,Parker2019}.

The crucial difference in the case of dilute sediment suspensions is particle
settling, which breaks the symmetry of the background stratification and
introduces a new parameter, the dimensionless fall velocity $v_s$.  In unsteady
flows, this value sets the size of vertical turbulent fluxes required to keep
particles in suspension and can determine whether the flow is laminar,
turbulent, or
intermittent~\citep{Cantero2009a,Cantero2012a,Shringarpure2012,Dutta2014,Langham2024}.
We study the effect of this parameter for various equilibrium solutions in a
shear-driven channel, finding that it profoundly affects both their ability to
transport sediment and the manner in which states are modulated by
stratification.  In brief, we show that vertical settling depresses the sediment
fluxes associated with the solutions in both extremal regimes of very high and
very low $v_s$. However, each of these regimes provides a separate mechanism
that desensitises states to the effects of density stratification, suggesting
that turbulent suspensions are most vulnerable to suppression at intermediate
$v_s$ values.  The results obtained dovetail with observations from DNS studies
of the same governing equations~\citep{Langham2024}, but benefit from the
possibility for complementary mathematical analysis, which we present alongside
numerical computations showing how the solutions adapt to parametric variations.

Following an exposition of the necessary technical background
in~\S\ref{sec:methods}, our study is divided into two main sections:
\S\ref{sec:passive scalar regime}, in which the particles are assumed to be
passively coupled to the flow and \S\ref{sec:stratification}, where the effect
of stratification is included.  In the passive regime (\S\ref{sec:passive scalar
regime}), solutions are converged numerically for a wide range of $v_s$ and
asymptotic expressions are derived for both the particle concentration fields
and resultant sediment fluxes, as settling becomes either very small, or very
large. On introducing stratification (\S\ref{sec:stratification}), we compute
solution curves parametrised by $\Ri_b$, unravel connections between different
states and examine how their structural dependence on $\Ri_b$ depends on $v_s$,
obtaining scaling laws for the existence of states with asymptotically low and
high settling velocity.  We conclude by discussing our findings
in~\S\ref{sec:discussion}.  This study also includes two appendices, both of
which help to guide the interpretation of the main findings. In the first, we
derive the expression for the evolution of mechanical and potential energy
within our model system which provides useful interpretation of the dynamical
process. In the second, we provide details of the associated linear stability
problem which may be used to initiate the search for new solutions and its
connections to known flow states.

\section{Methods and theoretical background}
\label{sec:methods}%
We consider plane Couette flow---that is, incompressible flow driven between
infinite parallel walls moving with equal and opposite velocities $\pm U_w
\vect{e}_x$, where $\{\vect{e}_x,\vect{e}_y,\vect{e}_z\}$ denotes an orthonormal
Cartesian frame with $\vect{e}_y$ oriented perpendicular to the plates.  We
shall commonly refer to these three coordinate directions as the ($x$)
\emph{streamwise}, ($y$) \emph{wall-normal} and ($z$) \emph{spanwise} flow
directions respectively.  To this canonical flow setup, we add a phase of
particles occupying a fraction $\psi(x,y,z,t)$ of the total volume and settling
under gravity with characteristic fall velocity $-V_s\vect{e}_y$.  The particles
are assumed to be monodisperse hard spheres of sufficiently small diameter that
they may be considered inertialess.  Furthermore, we suppose that their
concentration never exceeds $\sim\!\!1\%$ by volume, so that the physics of
inter-particle collisions may be safely neglected.  Finally, the density
difference between the particle--fluid mixture and the suspending fluid is
assumed to be small, so that the Boussinesq approximation may be applied.
The above considerations lead to the following governing equations for the flow
velocity $\vect{u}(x,y,z,t)$, pressure $p(x,y,z,t)$ and particle concentration
$c(x,y,z,t)$:
\begin{subequations}
\begin{gather}
    \frac{\partial\vect{u}}{\partial t}
    + \vect{u}\cdot\nabla\vect{u} 
    = -\nabla p
    + \frac{1}{\Rey}\nabla^2\vect{u} - \Ri_b c \vect{e}_y,\label{eq:governing 1}\\
    \nabla\cdot\vect{u} = 0,\label{eq:governing 2}\\
    \frac{\partial c}{\partial t} + (\vect{u} - v_s\vect{e}_y)\cdot \nabla c
    = \kappa \nabla^2 c,\label{eq:governing 3}
\end{gather}
\end{subequations}
where space and time have been made dimensionless with respect to the
inter-plate separation $h$ and the advective time scale $h/U_w$ and the
concentration $c$ has been normalised with respect to the total mass loading $M$,
i.e.\
\begin{equation}
    c = \frac{\psi}{M},\quad \textrm{where}\quad M = \frac{1}{|V|}\int_V \psi
    \,\mathrm{d}V.
    \label{eq:c normalisation}%
\end{equation}
The dimensionless parameters are the \emph{Reynolds number} $\Rey$, \emph{bulk
Richardson number} $\Ri_b$, \emph{settling velocity} $v_s$ and \emph{sediment
diffusivity} $\kappa$, given by
\begin{equation}
    \Rey = \frac{U_wh}{\nu}, \quad 
    \Ri_b = \frac{M(\varrho - 1)gh}{U_w^2}, \quad
    v_s = \frac{V_s}{U_w}, \quad
    \kappa = \frac{K}{U_wh},
\end{equation}
where $\nu$ is the kinematic viscosity of the suspending fluid, $\varrho$ is the
ratio of solid and fluid densities, $g$ is gravitational acceleration and $K$ is
a dimensional sediment diffusivity.  This latter parameter and its associated
term in Eq.~\eqref{eq:governing 3} is included in the governing equations to
homogenise over individual hydrodynamically mediated interactions between
particles that collectively act as an effective diffusive
process~\citep{Necker2002,Guazzelli2011,Shringarpure2012}. It is unrelated to
diffusive models arising from turbulence closures and does not require the
presence of turbulence to be active.  The class of governing equations described
by Eqs.~\eqref{eq:governing 1}--\eqref{eq:governing 3} has been used extensively
to model dilute suspensions, especially in the context of studying the structure
of turbidity currents \citep[see e.g.][]{Necker2002, Cantero2009a, Cantero2009b,
Cantero2012b,Cantero2012a,
Shringarpure2012,Salinas2021b,Salinas2021a}.

When $v_s = 0$, the system reduces to the standard Boussinesq
equations for flow that is density stratified, but not particle-laden. However, while studies of stratified
plane Couette impose a vertically-varying density gradient in the fluid phase,
via their boundary conditions or otherwise, here stratification arises
only due to the distribution of sediment, subject to vertical settling.
In addition to the standard no-slip conditions on the velocity for
plane Couette flow, we assume that there is no flux of concentration through
either of the bounding walls, applying
\begin{equation}
    \vect{u} = \pm \vect{e}_x
    \quad \textrm{and} \quad
    v_s c + \kappa \frac{\partial c}{\partial
    y} = 0
    \quad \textrm{at}~y=\pm 1.
    \label{eq:bcs}%
\end{equation}
This latter condition implies that the total mass of sediment in the channel is
conserved and leads to a homogeneous concentration field when $v_s = 0$. In
combination with the normalisation in~\eqref{eq:c normalisation},~\eqref{eq:bcs}
implies that the volume-averaged concentration is always unity.

Equations~\eqref{eq:governing 1}--\eqref{eq:governing 3}, together with the boundary
conditions~\eqref{eq:bcs} admit the following
one-dimensional base flow solution
\begin{subequations}
\begin{gather}
    \vect{u}_0(y) = y\vect{e}_x, \quad 
    p_0(y) = \frac{\Ri_b \exp(-v_s y / \kappa)}{\sinh(v_s/\kappa)}, \quad
    c_0(y) = \frac{v_s\exp(-v_s y / \kappa)}{\kappa\sinh(v_s/\kappa)}.
    \tag{\theequation\emph{a--c}}%
\end{gather}
    \label{eq:base flow}%
\end{subequations}
Note that the base sediment profile is controlled by the parameter combination
$v_s/\kappa$. This reflects the role that vertical diffusion plays in resisting
particle accumulation at the bottom wall. For this reason, it will often be
natural henceforth to measure the dimensionless settling velocity with respect
to the size of the dimensionless diffusion constant $\kappa$, even though we do
not systematically vary $\kappa$ in this study.

\subsection{Invariant solutions}
\label{sec:ecs}%
This article concerns steady state solutions (\emph{equilibria}) to
Eqs.~\eqref{eq:governing 1}--\eqref{eq:governing 3} and more generally,
streamwise \emph{travelling wave} states, which are solutions that are
steady in a reference frame that moves with a constant velocity $a \vect{e}_x$.
Together with other invariant flows, such as time-periodic
states, these solutions are commonly referred to as \emph{exact coherent
structures (ECS)}. They coexist with the laminar base flow and are generally
linearly unstable, though nonetheless dynamically
important~\citep{Kawahara2012,Graham2021}. Much is already
known about such solutions in homogeneous flows and it is useful to cover some
of the key points before proceeding.

In plane Couette flow without particles, the transition towards fully developed
turbulent flow begins with the emergence of oblique stripe patterns at $\Rey
\gtrsim 174$~\citep{Chantry2017}, which become sustained in finite experimental
realisations at around $\Rey \approx 360$~\citep{Tillmark1992}, in spite of the
linear stability of the laminar base flow. This coincides with the proliferation
of numerous equilibrium and travelling wave solutions, which are primarily born
in saddle-node bifurcations that start to populate the state space of solutions
at around the critical $\Rey$ for the onset of stripe
formation~\citep{Gibson2009,Reetz2019,Ahmed2020}.  Each such bifurcation creates
two families of states: an \emph{upper} and \emph{lower} branch, with the lower
branch solutions lying closer to the base flow with respect to commonly used
metrics (e.g.\ $L^2$-norm).  Various lower branch states have been shown
directly to play a role in mediating turbulence
transition~\citep{Wang2007,Gibson2008,Schneider2008,Reetz2019}, while the more
energetic upper branch solutions populate the region of state space visited by
unsteady flow simulations~\citep{Gibson2008,Gibson2009}.

Almost all known unstable equilibrium states emerge from the
following tripartite interaction: streamwise-invariant \emph{rolls} with
velocity field $[0, V(y,z), W(y,z)]$ redistribute the base flow profile
[Eq.~(\ref{eq:base flow}\emph{a})], giving rise to \emph{streaks}, whose
spanwise-inflectional velocity profile $[U(y,z), 0, 0]$ is linearly unstable to
streamwise modes.  The resultant waves are in turn responsible for forcing the
rolls. This is known as the \emph{self-sustaining process~(SSP)} \citep{Waleffe1997}.
In the asymptotic regime of high Reynolds number, the relative sizes of these
components is captured by the \emph{vortex-wave interaction~(VWI)} theory of
\cite{Hall1991}.  The flow fields take the form
\begin{subequations}
\begin{gather}
    u = U(y,z) + \ldots + \delta'(\Rey)\left[\widetilde{u}(y,z)\mathrm{e}^{\im k(x-at)} +
    \mathrm{c.c.}\right] + \ldots,\\
    v = \Rey^{-1} V(y,z) + \ldots + \delta'(\Rey)\left[\widetilde{v}(y,z)\mathrm{e}^{\im k(x -
    at)} +
    \mathrm{c.c.}\right] + \ldots, \\
    w = \Rey^{-1} W(y,z) + \ldots + \delta'(\Rey)\left[\widetilde{w}(y,z)\mathrm{e}^{\im k(x-at)} +
    \mathrm{c.c.}\right] + \ldots, \\
    p = \Rey^{-2} P(y,z) + \ldots + \delta'(\Rey)\left[\widetilde{p}(y,z)\mathrm{e}^{\im k(x-at)} +
    \mathrm{c.c.}\right] + \ldots,
\end{gather}
    \label{eq:vwi}%
\end{subequations}
where the latter terms of these expansions capture the streamwise wave component
with nonzero wavenumber $k$, speed $a$ and amplitude whose magnitude (w.r.t.\
$\Rey$) is given by $\delta'(\Rey) = \Rey^{-7/6}$, save for within in an
$O(\Rey^{-1/3})$ thickness critical layer, where $U=a$ and $\delta'(\Rey) =
\Rey^{-5/6}$. In the limit $\Ri_b \to 0$, where flow velocity is uncoupled to
the sediment concentration, the equilibrium and travelling wave solutions
studied herein are finite-$\Rey$ analogues of these VWI
states~\citep{Hall2010,Deguchi2014}. Though $\Rey$ is fixed at a relatively
modest value ($400$) for this study, the SSP/VWI framework remains useful in
these regimes and provides a means for understanding how states develop as $\Rey
\to \infty$. Moreover, it has previously been used to analyse the effect of
stratification on states~\citep{Eaves2015,Olvera2017,Deguchi2017,Langham2020}.

It is frequently useful to extract just the streak and roll components of a
flow structure. Denoting the streamwise period of a solution by $L_x = 2\pi/k$, we
define the streamwise average $\bar{f}$ of a given field $f$, by
\begin{equation}
    \bar{f}(y,z) = \frac{1}{\! L_x} \int_0^{L_x} \! f(x,y,z) \,\mathrm{d}x.
\end{equation}
Likewise, spanwise averages shall also prove useful.
Assuming an $L_z$-periodic state, we employ the following notation for these
operations:
\begin{equation}
    \langle f\rangle(x,y) = 
    \frac{1}{L_z}\int_0^{L_z} 
    \!
    f(x,y,z)
    \,\mathrm{d}z.
\end{equation}
Simultaneous averages in both wall-parallel directions, which convey
information about mean vertical profiles, are therefore written as
$\langle \bar{f}\rangle(y)$. (In each of these cases, the coordinate dependence will
typically be omitted for brevity.)

All equilibrium and travelling wave states must match the steady energy
input~$I$
by the boundary shear with corresponding energetic losses.  In homogeneous flow,
the only mechanism available is the dissipation~$D$ from internal viscous stresses.
However, when relatively dense particles are present, the work done by the
velocity field to maintain the suspension must also be considered. 
The (dimensionless) instantaneous total energy density is
\begin{equation}
    E_{\textit{tot}} = \frac{1}{2}\int_{-1}^1 \left\langle
    \overline{
    \tfrac{1}{2}\vect{u}\cdot\vect{u}
    +
    \Ri_b yc
    }
    \right\rangle \,\mathrm{d}y.
    \label{eq:energy density}%
\end{equation}
Using details provided in Appendix~\ref{appendix:energy}, it may be verified
that under steady state conditions, the balance 
\begin{equation}
    I = D + \frac{\Rey\Ri_b}{2}\!\int_{-1}^{1}\langle \overline{vc} \rangle\,\mathrm{d}y,
    \label{eq:energy balance}%
\end{equation}
is satisfied, where (using a colon to denote the Frobenius inner product of
matrices) we define the input and dissipation
\begin{subequations}
\begin{equation}
    I =
        \frac{\mathrm{d} \langle \bar{u}\rangle}{\mathrm{d} y}\bigg|_{y=1},
    \quad
    D =
    \frac{1}{2}
    \int_{-1}^{1}
    \langle\overline{\nabla \vect{u}:\nabla\vect{u}}\rangle\,\mathrm{d}y,
    \tag{\theequation\emph{a,b}}%
\end{equation}
    \label{eq:I and D}%
\end{subequations}
normalised such that
%
the mean wall stress $\tau_y = I$ is unity for laminar flow.
To characterise different states in the forthcoming sections, we use
$\hat{\tau}_y$, the
fraction of wall stress corresponding to the deviation from laminar flow,
defined by $\hat{\tau}_y = \tau_y - 1$.
Other quantities corresponding to the perturbation from the base
state will also be adorned with hats. For example, the velocity fields
\begin{equation}
    \hat{u} = u - y, ~~ \hat{v} = v, ~~ \hat{w} = w, ~~ \hat{c} = c - c_0,
    \label{eq:perturbations}%
\end{equation}
with $c_0$ as defined in Eq.~(\ref{eq:base flow}\emph{c}).

\subsection{Numerical methods}
Equilibrium and travelling wave ECS were computed using a version of the
Channelflow~2.0 software package~\citep{Channelflow} that was adapted to
integrate Eqs.~\eqref{eq:governing 1}--\eqref{eq:governing 3} with the
corresponding boundary conditions [Eq.~\eqref{eq:bcs}] and used previously to
investigate the system using direct numerical simulations~\citep{Langham2024}.
States were converged in the domain $(x,y,z) \in[0, L_x] \times [-1, 1] \times
[0, L_z]$, using a pseudo-spectral discretisation comprising $M_x$ and $M_z$
Fourier modes in $x$ and $z$ respectively (thereby enforcing periodicity in
these directions) and $M_y$ Chebyshev polynomials in $y$.  Typical resolutions
were $(M_x, M_y, M_z) = (48, 35, 48)$, with $16$ de-aliased modes in $x$~\&~$z$.
For solutions that developed narrow boundary layers in the wall-normal
direction, $M_y$ was increased up to~$181$.  Throughout the computations, we
verified the insensitivity of various states to higher resolution by manually
checking their fields and bulk statistics. For the solutions listed below in
table~\ref{tab:tw properties}, resolution was increased until the final
significant digits in the reported quantities ceased to vary.  Computations of
individual ECS involve searching for zeros of the time-forward integration map
of the governing equations, using a Newton-hookstep
algorithm~\citep{Dennis1996}, together with a matrix-free linear solver (GMRES)
and a suitable initial guess. These are standard techniques that are documented
in detail elsewhere~\citep{Viswanath2007,Gibson2009}. Initially, suitable guesses
for the Newton solver were obtained from an existing library of known ECS in
homogeneous plane Couette flow~\citep{Gibson2009}, with $L_x = 2\pi/1.14$, $L_z
= 2\pi / 2.5$. Corresponding guesses for the sediment field were constructed by
time stepping Eq.~\eqref{eq:governing 3} with respect to a fixed velocity field
given by the homogeneous ECS and starting from a constant concentration profile
with unit mass loading, until $c(\vect{x},t)$ relaxed to steady state.
Once these initial solutions were converged, additional states were obtained by
using arclength continuation in one or more of the system parameters ($\Rey$,
$\Ri_b$, $v_s$ and $\kappa$) to generate guesses for the Newton solver along
solution branches.
Eigenvalue computations for table~\ref{tab:tw properties} were performed via
Arnoldi iteration.

\subsection{Parameter selection and physical interpretation}
\label{sec:parameters}%
To keep the numerical component of this study manageable, we make our
primary focus the effects of $v_s$ and $\Ri_b$, which characterise the strength
of vertical particle settling and buoyancy forces respectively.  We assume
throughout that particles settle in the negative $y$-direction, i.e.\ $v_s>0$.
This sign convention implies that $\Ri_b < 0$ is unphysical,
since it means that particle settling opposes the direction of buoyancy.
Nevertheless, it is useful on a technical level to consider such states
since, through parametric continuation (see~\S\ref{sec:vk 1}), they make connections with solutions in physical regimes.
Note that the restriction $v_s>0$ is
applied without loss of generality, since any solution of the governing
equations~\eqref{eq:governing 1}--\eqref{eq:governing 3}, corresponds to an
equivalent state for which the sediment buoyancy takes the opposite sign, under
the transformations
\begin{gather}
    \Ri_b \mapsto -\Ri_b, ~~ v_s \mapsto -v_s,~~
    [u,v,w,p,c](x,y,z,t) \mapsto [-u, -v, w, p, c](-x, -y, z, t).
    \label{eq:settling symmetry}%
\end{gather}
The Reynolds number and sediment diffusivities are fixed at $\Rey = 400$ and
$\kappa = 2.5\times 10^{-3}$ respectively, except in~\S\ref{sec:passive low vs},
where higher $\Rey$ values are briefly investigated and $\kappa$ is adjusted in
tandem, in order to maintain unit \emph{Schmidt number} $\Sc = \nu / K = 1$.  We
favour the choice of $(\Rey,Sc)=(400,1)$ for this initial study, chiefly because
it helps us to compare the effect of sediment on this system against prior
studies in the homogeneous and density stratified cases which have used these
values~\citep{Gibson2009,Olvera2017,Ahmed2020}.  Moreover, note also that $\Sc =
1$ places our system within the range of sediment diffusivity estimates for fine
sand particles in water~\citep{Shringarpure2012} and represents the default
choice for turbulent
simulations~\citep{Necker2002,Cantero2009a,Cantero2009b,Cantero2012b,Cantero2012a,Shringarpure2012,Salinas2021b,Salinas2021a}.
Furthermore, since the effects of $\Rey$ and $\kappa$ have been studied for the
case of density stratified invariant solutions in plane Couette
flow~\citep{Olvera2017,Langham2020}, it is possible to gain insight into their
effects by considering this prior work, in conjunction with the theoretical
contributions in the following sections.  

The model system~\eqref{eq:governing 1}--\eqref{eq:governing 3} has been
used to investigate flows with direct geophysical
applications~\citep[e.g.][amongst others]{Shringarpure2012,Salinas2021b},
including the turbulent dynamics of fine sand of diameters up to $\sim
100\mu\mathrm{m}$ and density $\sim 2500\mathrm{kg\,m}^{-3}$, which settles
through quiescent water with velocity $\sim
10^{-2}\mathrm{m}\,\mathrm{s}^{-1}$.
Larger particles are influenced by non-negligible inertial effects and thus are
not well represented by this modelling framework.
In environmental
flows, where $\Rey$ is typically far higher than numerical simulations can
achieve, this settling velocity is low enough that particles spend considerable
time in suspension, yet sufficiently high that mean particle concentrations
are much larger towards the flow base.  The particular form of the
concentration profile is contingent on the size of the characteristic turbulent
shear velocity, but can be strong enough to produce stratifications that
suppress turbulence, even at low
concentrations~\citep{Villaret1991,Barenblatt1996}.
Indeed, it will be shown later, in \S\ref{sec:vk 1} that quite small values of 
$\Ri_b$ can be sufficient to disrupt significantly the invariant states studied 
herein, echoing observations of turbulence simulations that show suppression and
laminarisation for $\Ri_b\sim0.01$~\citep{Langham2024}.

Due to the mismatch between the moderate Reynolds number of this study and flows
at field scale, there are two potential interpretations of the work presented
below. In the most direct sense, the ECS capture the interaction between
sediment and vortical structures at the smallest scales, embedded within the
sheared wall region of more spatiotemporally complex flows.  In this way, the
results present a first investigation of the effects of exact coherent flow
states on sediment and their disruption by the induced density gradients.
Alternatively, they can be viewed as a simplified caricature that captures the
interplay between the processes of settling, shear and vertical sediment flux
that are present at all scales in turbulent suspensions.  This mirrors the
approach of DNS studies, which can provide worthwhile qualitative insight beyond
the range of $\Rey$ that may be feasibly studied.
If this latter interpretation is adopted, $v_s$ and $\Ri_b$ should be matched to
give a concentration profile and density stratification that represents a given
flow of interest.
In this regard, then cases with $v_s/\kappa \gg 1$, which lead to sediment
concentrations that strongly diminish as $y$ increases (as may be verified by
reading on, or considering~ Eq.~(\ref{eq:base flow}\emph{c})) correspond to the
upper range of sediments that can be modelled by our system. Smaller particles,
such as very fine sands, silts and aerosols settle more gradually, leading to
weaker concentration gradients, which may be captured with progressively lower
$v_s/\kappa$, ultimately culminating in the $v_s/\kappa \ll 1$ regime, where
sediment distribution is near-homogeneous.  To account for this, our analysis
covers both extremal regimes and likewise varies $\Ri_b$ until nontrivial
solutions cease to exist.

\subsection{Reference states in homogeneous flow}
In order to study how invariant solutions adapt to the presence of settling
particles, we use known solutions in unstratified plane Couette flow as initial
reference cases.  A convenient database is provided by the study
of~\cite{Gibson2009} who converged $13$ equilibria at $\Rey = 400$, naming them
$EQ_1$--$EQ_{13}$.  Though each of these states has particular properties, they
all share the same fundamental roll--streak--wave structure dictated by the
SSP/VWI asymptotic theory [see \S\ref{sec:ecs} and Eqs.~(\ref{eq:vwi}\emph{a--d})]. Consequently, we expect each family of solutions to
react to parametric variations in a qualitatively similar way. Therefore, we
focus our attention primarily on the simplest of these states: $EQ_1$ and its
upper branch counterpart~$EQ_2$.

\begin{figure}
    \includegraphics[width=\columnwidth]{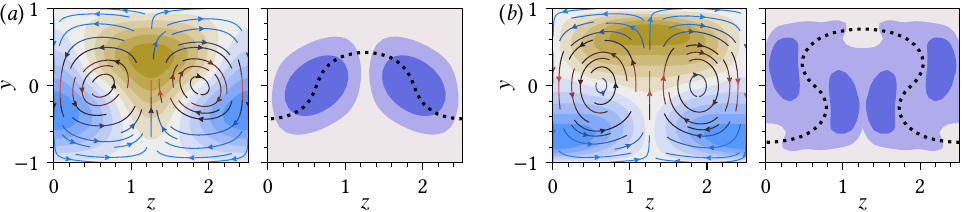}
    \caption{Example lower and upper branch equilibrium solutions at $\Rey =
    400$ and $\Ri_b=0$:
    (\emph{a})~$EQ_1$ and (\emph{b})~$EQ_2$.  The left-hand panels in each pair
    show filled contours of the streamwise-invariant component of the streamwise
    velocity perturbation to the base state, $\bar{u}(y,z) - y$, coloured
    from brown to blue using $10$ equispaced intervals centred around zero (white)
    and extending to $\pm 0.45$ for $EQ_1$ and $\pm 0.76$ for $EQ_2$.
    Streamlines show the cross-stream rolls
    $[\bar{v},\bar{w}](y,z)$, coloured according to $|(\bar{v},\bar{w})|$, whose
    maximum values are $0.018$ for $EQ_1$ and $0.098$ for $EQ_2$ (2~s.f.).
    Right-hand panels show filled contours equispaced within $[0,0.052]$ for
    $EQ_1$ and $[0,0.13]$ for $EQ_2$, delineating the
    average amplitude of streamwise variation in $u$, given by $\overline{|u -
    \bar{u}|}(y,z)$.
    The critical layers (where $\bar{u} = 0$) are plotted in dotted black.
    }
    \label{fig:eq1}
\end{figure}

These solutions were originally discovered by~\cite{Nagata1990}. $EQ_1$ consists of two
streamwise-aligned counter-rotating rolls, depicted in
figure~\ref{fig:eq1}(\emph{a}), which act to advect fluid with $u < 0$ upwards to
create the low-speed streak coloured in brown and likewise draw $u > 0$ patches
downward, creating the blue high-speed streak.
The streamwise dependence in the velocity fields is organised around the
critical layer. Variations in $u$ are visualised in the second panel. They
diminish and localise towards the critical layer (dotted black) as $\Rey$
increases, in accordance with the SSP/VWI framework outlined in \S\ref{sec:ecs}.
[The
convergence w.r.t.\ $\Rey$ is not shown here, but see the work
of~\cite{Hall2010,Deguchi2014}.] This underpins our decision to primarily
visualise states using streamwise-averaged contour plots, despite the fact that
ECS can exhibit noticeable streamwise waviness at $\Rey=400$ \citep[see figure~4
of][]{Gibson2009}.  In doing so, the essential roll--streak structures that come
to dominate solutions as $\Rey \to \infty$ are highlighted.
Figure~\ref{fig:eq1}(\emph{b}) shows $EQ_2$, the corresponding upper branch
state, which shares the same essential structure, but features higher stress at
the walls and more energetic rolls.

The velocity fields of $EQ_1$ and $EQ_2$ are invariant under the group of
symmetry transformations generated by the following two elements:
\begin{subequations}
\begin{gather}
    \mathcal{R}:[u,v,w](x,y,z) \mapsto
    [-u,-v,w](-x+L_x/2,-y,z+L_z/2),\label{eq:shift-and-rotate}\\
    \mathcal{S}:[u,v,w](x,y,z) \mapsto [u,v,-w](x+L_x/2,y,-z).
\end{gather}
\end{subequations}
The first of these is a streamwise and spanwise shift by half of the computational cell,
followed by a rotation of $\pi$ around the $z$ axis, which flips the signs of
$u$ and $v$. 
This symmetry enforces zero bulk streamwise velocity $\int_{-1}^{1}\langle
\bar{u}\rangle\,\mathrm{d}y=0$, as well as preventing any
streamwise-dependent structures from moving in the $x$ direction (since this
would break the symmetry), thereby enforcing $a=0$.
Similarly, $\mathcal{S}$ denotes a streamwise half-cell shift,
followed by spanwise reflection, which flips the sign of $w$ and anchors the
solution in place along the $z$ direction.

\section{Passive scalar regime ($\Ri_b=0$)}
\label{sec:passive scalar regime}%
\begin{figure}
    \includegraphics[width=\columnwidth]{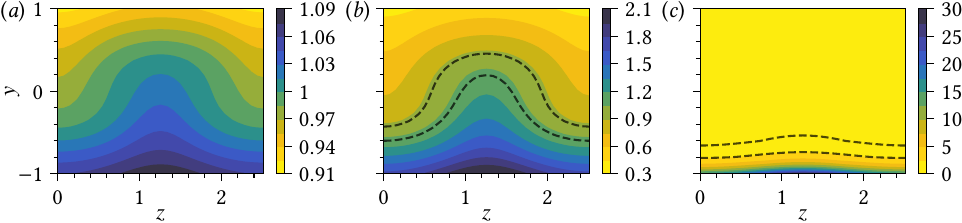}
    \caption{Contours of $\bar{c}(y,z)$ for $EQ_1$ at different settling
    velocities: (\emph{a})~$v_s/\kappa = 0.1$, (\emph{b})~$v_s/\kappa = 1$ and
    (\emph{c})~$v_s/\kappa = 10$.  For comparison across panels, in
    (\emph{b},\emph{c}), two dashed black contours are located at the
    corresponding maximum and minimum values of $\bar{c}$ in panels
    (\emph{a},\emph{b}) respectively.
    }
    \label{fig:cfields}
\end{figure}

To study how these states accommodate a phase of settling particles,
we begin by computing concentration fields 
in the limiting case $\Ri_b = 0$. Since we retain the effects
of particle settling, this regime is best understood as the physical limit where
sediment concentrations are too dilute to affect the flow physics via buoyancy.
In figure~\ref{fig:cfields}, we show three equilibrium concentration fields
converged from $EQ_1$ with different settling velocities: $v_s/\kappa = 0.1$, $1$
and $10$.
In each case, the converged sediment field possesses the same
`shift-and-reflect' $\mathcal{S}$ symmetry as the $u$ field, but violates 
$\mathcal{R}$, since nonzero particle settling prevents there being an equal
distribution of sediment in the upper and lower halves of the channel (modulo a
streamwise shift).
For the lowest settling velocity
[figure~\ref{fig:cfields}(\emph{a})], the concentration is close to being
spatially uniform,
deviating no more than $10\%$ from unity. Nevertheless, the effect of the
velocity structure in shaping the equilibrium concentration field is clear: more
concentrated patches are drawn up from the bottom wall by the rolls at $z = L_z/2$
and less concentrated patches likewise emerge at $z = 0$, where $\bar{v}<0$ [see
figure~\ref{fig:eq1}(\emph{a})]. Increasing the settling velocity so that
$v_s/\kappa = 1$, leads to greater vertical stratification in the $\bar{c}$
field [figure~\ref{fig:cfields}(\emph{b})], with preferential accumulation of
concentration at the bottom wall.  When $v_s$ is far greater than $\kappa$, this
latter trend exaggerates. At $v_s/\kappa = 10$
[figure~\ref{fig:cfields}(\emph{c})], we observe a strongly concentrated boundary
layer in the vicinity of $y = -1$, beneath a dilute upper region.  Though the
effect is barely discernible on a linear scale, we note that even at this
highest settling velocity, the rolls act to vertically redistribute the
concentration field in accordance with the sign of $\bar{v}$.

Both the extremal regimes, $v_s/\kappa \ll 1$ and $v_s/\kappa \gg 1$, are
amenable to asymptotic analyses that explain the resultant structures of the
concentration fields in response to the imposed flow. 
Since the velocity and pressure fields are unaffected by
$c$ in the passive limit, Eq.~\eqref{eq:governing 3} decouples from
Eqs.~\eqref{eq:governing 1} and~\eqref{eq:governing 2} and may be considered in
isolation.  Given a travelling wave solution with wave velocity $\vect{a} = a
\vect{e}_x $, we write Eq.~\eqref{eq:governing 3} in terms of perturbations to
the base flow using the notation of~\eqref{eq:perturbations}, to give
\begin{equation}
    (y + \hat{u} - a)\frac{\partial \hat{c}}{\partial x}
    + \hat{v}\frac{\partial c_0}{\partial y}
    + (\hat{v} - v_s)\frac{\partial \hat{c}}{\partial y}
    + \hat{w}\frac{\partial \hat{c}}{\partial z}
    =
    \kappa \nabla^2 \hat{c}.
    \label{eq:c eqn}%
\end{equation}
In the case of the equilibria $EQ_1$ and $EQ_2$, $a = 0$, but more generally
the exact coherent state could exhibit a non-vanishing wave velocity.

\subsection{Low settling velocity $(v_s/\kappa \ll 1)$}
\label{sec:passive low vs}%
\begin{figure}
    \includegraphics[width=\columnwidth]{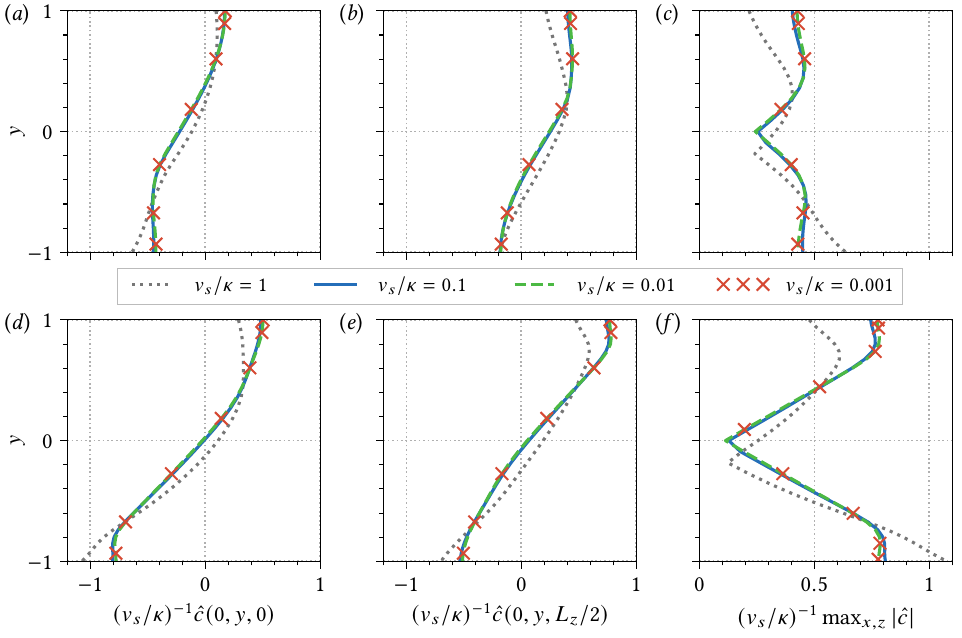}
    \caption{%
    Rescaled perturbations to the concentration field,
    in the regime $v_s/\kappa \ll 1$.
    The top row of panels reports data for $EQ_1$.
    (\emph{a,b})~Vertical profiles of $\hat{c} = c - c_0$ at $x = 0$, $z = $
    (\emph{a})~$0$ and (\emph{b})~$L_z/2$,
    rescaled by $v_s/\kappa$.
    (\emph{c})~Vertical profiles of the absolute maximum of $\hat{c}$ over the
    horizontal coordinates $x$ and $z$, rescaled by $v_s/\kappa$.
    Panels~(\emph{d}--\emph{f}) along the bottom row report the corresponding
    data for the upper branch state,~$EQ_2$.
    }
    \label{fig:lowvs_cpert}
\end{figure}

When $v_s = 0$, $c_0 \equiv 1$ and Eq.~\eqref{eq:c eqn} admits the trivial
homogeneous perturbation $\hat{c} \equiv 0$, regardless of the flow state.
Likewise, as shown in figure~\ref{fig:cfields}(\emph{a}), when $v_s/\kappa$ is
small but nonzero, the concentration field is a small correction to this
homogeneous state.  We expand the concentration field as $c = c_0(y) + \epsilon
c_1(x,y,z) + \ldots$, where $\epsilon$ is a positive order parameter assumed to
be small, relative to the fields $c_0$ and $c_1$. After substituting into
Eq.~\eqref{eq:governing 3}, neglecting $O(\epsilon^2)$ terms and using
Eq.~(\ref{eq:base flow}\emph{c}), we obtain
\begin{equation}
    \epsilon(y+\hat{u} - a)\frac{\partial c_1}{\partial x}
    + \hat{v} \left(
    -\frac{v_s}{\kappa}c_0
        + \epsilon \frac{\partial c_1}{\partial y}
    \right)
    -\epsilon v_s \frac{\partial c_1}{\partial y}
    + \epsilon\hat{w}\frac{\partial c_1}{\partial z}
    =
    \epsilon \kappa \nabla^2 c_1.
    \label{eq:lowpr eqn full}%
\end{equation}
Provided $\Sc \sim 1$, the settling term $\epsilon v_s\partial c_1/\partial y$ can be considered negligible, on the
grounds that $v_s \ll \kappa \ll 1$, so $\epsilon v_s \ll \epsilon$.  The size
of the remaining $O(\epsilon)$ terms in Eq.~\eqref{eq:lowpr eqn full} must match
the vertical advection of the base concentration profile, so $\epsilon =
v_s/\kappa$. 
This leads to 
\begin{equation}
    (y + \hat{u} - a) \frac{\partial c_1}{\partial x}
    + \hat{v}\frac{\partial c_1}{\partial y}
    + \hat{w}\frac{\partial c_1}{\partial z}
    = \hat{v} + \kappa \nabla^2 c_1,
    \label{eq:lowvs c1 eq}%
\end{equation}
where we have used the fact that $c_0 = 1 + O(\epsilon)$ when $v_s/\kappa \ll
1$.  
We demonstrate the inferred scaling regime in figure~\ref{fig:lowvs_cpert}, for
both the solutions $EQ_1$ and $EQ_2$.
For example, in figure~\ref{fig:lowvs_cpert}(\emph{a},\emph{d}), plotted 
profiles of $\hat{c}$ at
$(x,z) = (0,0)$ collapse under rescaling by $v_s/\kappa$, for
$v_s/\kappa = 10^{-1}$, $10^{-2}$ and $10^{-3}$. 
For reference, we also show the
corresponding profiles for $v_s/\kappa=1$ (dotted grey), which do not collapse
onto the limiting curves, but nevertheless lie close by. Profiles taken at other
points in the domain exhibit similar collapse 
-- figures~\ref{fig:lowvs_cpert}(\emph{b},\emph{e}) demonstrate the case
$(x,z)=(0,L_z/2)$.
To summarise
this, in figures~\ref{fig:lowvs_cpert}(\emph{c},\emph{f}), we plot
$(v_s/\kappa)^{-1}\max_{x,z}|\hat{c}|$ for the same $v_s/\kappa$ values, which
attain a limiting profile for $v_s/\kappa \lesssim 10^{-1}$.  

\begin{figure}
    \includegraphics[width=\columnwidth]{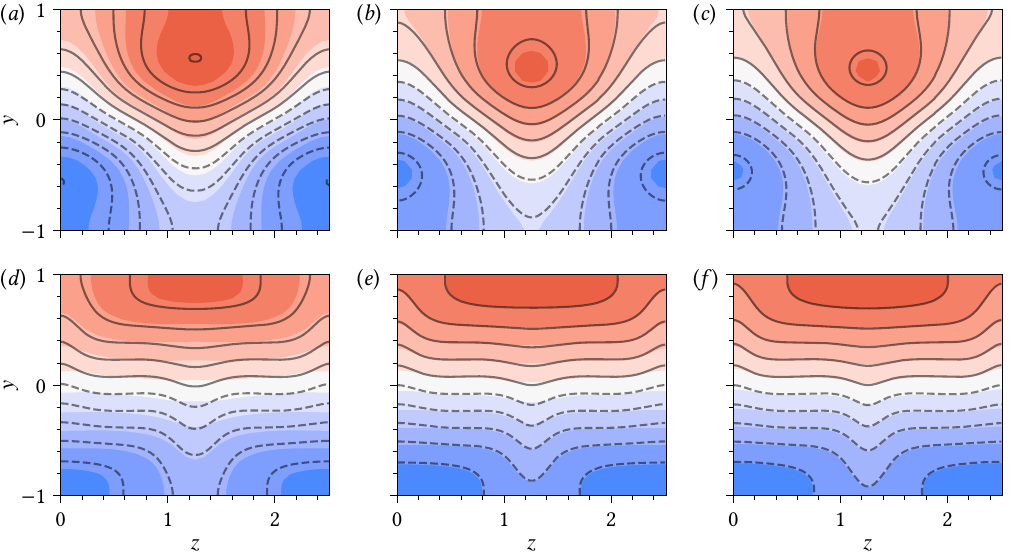}
    \caption{Comparison between the ECS concentration perturbations and 
    asymptotic solutions to
    Eq.~\eqref{eq:lowvs final} in the regime $v_s/\kappa\ll 1$, for increasing $\Rey$.
    (\emph{a}--\emph{c})~Filled contours from blue (low) to red (high)
    show $\bar{c} - c_0$ for $EQ_1$ at
    $v_s/\kappa = 10^{-2}$, equally spaced between 
    $\pm 0.5 v_s/\kappa$
    and centred at zero (white).  Isolines of $(v_s/\kappa)\bar{c}_1$ are
    overlaid at the same contour values with solid and dashed lines representing
    positive and negative values respectively.
    The Reynolds numbers are $\Rey = $ (\emph{a})~$400$, (\emph{b})~$1200$
    and (\emph{c})~$2000$.
    Panels~(\emph{d}--\emph{f}) present the corresponding data for $EQ_2$ in the
    same style, except the contours are taken at $\pm 0.9 v_s/\kappa$.
    }
    \label{fig:lowvs higher re}
\end{figure}

Further insight into the asymptotic structure may be obtained by appealing to
the SSP/VWI scalings for the velocity fields.  
Noting that these are uncoupled
to $c$, we substitute them~[Eqs.~(\ref{eq:vwi}\emph{a--c})] into
Eq.~\eqref{eq:lowvs c1 eq}. 
Away from the asymptotically shrinking critical layer,
it may be deduced that $\partial
c_1 / \partial x \to 0$ as $\Rey \to \infty$.
The resulting equation may be averaged to obtain the balance that
sets $\bar{c}_1$,
\begin{equation}
    V\frac{\partial \bar{c}_1}{\partial
    y} + W\frac{\partial \bar{c}_1}{\partial z}
    = V + \frac{1}{\Sc} \left(
    \frac{\partial^2 \bar{c}_1}{\partial y^2} +
    \frac{\partial^2 \bar{c}_1}{\partial z^2}
    \right)\!,
    \label{eq:lowvs final}%
\end{equation}
where $\Rey^{-1}(V, W) = (\bar{v}, \bar{w})$ and subject to the boundary
conditions $\partial \bar{c}_1/\partial y=0$ at $y = \pm 1$.
We solve this numerically, 
using a pseudospectral representation of $\bar{c}_1(y,z)$ with Fourier modes in
$z$ and Chebyshev polynomials in $y$. The velocity fields
$\bar{v}$ and $\bar{w}$ are provided by the known solutions for either $EQ_1$ 
or $EQ_2$.
This leads to
asymptotic predictions for the concentration perturbations of each state
at high $\Rey$ and low $v_s/\kappa$, which
are compared with the corresponding 
$c - c_0$ fields
for the full states at $v_s/\kappa
= 10^{-2}$ in figure~\ref{fig:lowvs higher re}.
The $\Rey = 400$ cases are given in panels~(\emph{a},\emph{d}) and exhibit
reasonable agreement.
Sediment concentration is elevated with respect to the base profile towards the upper
wall and diminished towards the lower wall.  By referring to figure~\ref{fig:eq1},
we note that these trends are accentuated according to the locations of the
rolls, which redistribute particles to form dilute and concentrated patches
around the points $(y,z) = (-1,0) \equiv (-1,L_z)$ and $(1,L_z/2)$
respectively.
As $\Rey$ is increased, to $1200$ in figures~\ref{fig:lowvs higher re}(\emph{b,e})
and then to $2000$ in figures~\ref{fig:lowvs higher re}(\emph{c,f}), the
asymptotic solutions and full ECS states converge to one another.

\subsection{High settling velocity $(v_s/\kappa \gg 1)$}
\label{sec:high vs}%
When the settling velocity is relatively large, a layer of sediment forms at the bottom wall
[see figure~\ref{fig:cfields}(\emph{c})].  Increasing $v_s$ concentrates the boundary
layer, whose characteristic width~$\varepsilon$ (to be determined below) appears to shrink to zero in the
limit $v_s\to\infty$.  Therefore, we analyse this region by rescaling the
vertical coordinate at the wall, defining $Y = (y + 1) / \varepsilon$.
The concentration equation then becomes
\begin{equation}
    \frac{\partial^2 \hat{c}}{\partial Y^2}
    +
    \varepsilon \frac{v_s}{\kappa}\frac{\partial \hat{c}}{\partial Y}
    =
    \frac{\varepsilon^2}{\kappa}(-1 - a + \hat{u} + \varepsilon Y)
    \frac{\partial \hat{c}}{\partial x}
    + \frac{\varepsilon}{\kappa} \hat{v}\left(
    \frac{\partial c_0}{\partial Y}+
    \frac{\partial \hat{c}}{\partial Y}
    \right)
    +\frac{\varepsilon^2}{\kappa}\hat{w}\frac{\partial \hat{c}}{\partial z}
    -\varepsilon^2\left(
    \frac{\partial^2 \hat{c}}{\partial x^2}+
    \frac{\partial^2 \hat{c}}{\partial z^2}
    \right)\!.
    \label{eq:high vs c eqn}%
\end{equation}
In order for nontrivial solutions to exist as $v_s/\kappa$ becomes high,
$\varepsilon = \kappa / v_s$. (Although $\partial c_0/\partial Y$ also depends
on $v_s/\kappa$, it will become clear shortly that this term is subdominant.)
The only terms that survive to
leading order as $\varepsilon\to 0$ are vertical settling and vertical diffusion, 
\begin{equation}
    \frac{\partial^2 \hat{c}}{\partial Y^2}
    +
    \frac{\partial \hat{c}}{\partial Y} = 0.
\end{equation}
This equation has general solution $\hat{c}(x,Y,z) = A(x,z) \mathrm{e}^{-Y} +
B(x,z)$.  The no-flux boundary at $Y=0$ implies that $B \equiv 0$.  The
amplitude $A(x,z)$ is dictated the right-hand side terms of Eq.~\eqref{eq:high
vs c eqn}, which are present at higher order.  The magnitudes of these terms are
dependent on the velocity field.  For sufficiently small $\varepsilon$, the VWI
critical layer (whose location is fixed) must lie outside the boundary layer.
Consequently, the velocity structure is streamwise independent within this
region and is given by the streak and roll components of
Eqs.~(\ref{eq:vwi}\emph{a--c}).
These may be Taylor expanded at the wall to infer that
\begin{subequations}
\begin{gather}
    \hat{u}(Y,z) = \varepsilon u_1(z) Y + \ldots,\\
    \hat{v}(Y,z) = 
    \Rey^{-1}\varepsilon^2 v_2(z) Y^2 + \ldots,\\
    \hat{w}(Y,z) = \Rey^{-1}\varepsilon w_1(z) Y+\ldots,
\end{gather}
    \label{eq:vwc expansions}%
\end{subequations}
where $u_1$, $v_2$ and $w_1$ are unknown functions that are $O(1)$ with respect
to both $\varepsilon$ and $\Rey$.
The quadratic dependence of the wall-normal velocity field follows from the
incompressibility of the flow.
Substitution of these expressions into Eq.~\eqref{eq:high vs c eqn}, along with the
asymptotic formula for the base concentration field $c_0(Y)=
2\varepsilon^{-1}\mathrm{e}^{-Y} + \ldots$, implies that the right-hand side
terms in Eq.~\eqref{eq:c eqn} enter at $O(\varepsilon^2)$.
Therefore, the concentration perturbation has the asymptotic form
\begin{gather}
    \hat{c}(x,Y,z) = A(x,z) \mathrm{e}^{-Y} + \varepsilon^2 c_2(x,Y,z) + \ldots,
    \label{eq:c expan}%
\end{gather}
where $c_2$ is an arbitrary $O(1)$ field.
Although streamwise dependence has been retained in this expansion, in practice the
concentration is streamwise invariant to leading order,
because its structure is controlled by the $x$-invariant velocity
fields.
Consequently, to effectively characterise $\hat{c}$ at leading order, it is
enough to identify $\bar{A}(z)$, which may be obtained by substituting
Eq.~\eqref{eq:c expan} into
Eq.~\eqref{eq:high vs c eqn} (along with the expressions for the velocities and $c_0$)
and streamwise averaging.
At $O(\varepsilon^2)$, this leads to
\begin{equation}
\frac{\partial^2 \bar c_2}{\partial Y^2}+
\frac{\partial \bar c_2}{\partial Y}    
    =-2\Sc\, v_2(z) Y^2 \mathrm{e}^{-Y} - \bar A''(z)\mathrm{e}^{-Y}.
\end{equation}
Integrating this equation in the wall-normal direction between $0$ and $Y$ gives
\begin{equation}
\frac{\partial \bar c_2}{\partial Y^2}+
\bar c_2
    =2\Sc\, v_2(z) \mathrm{e}^{-Y}(Y^2 + 2Y + 2) + \bar{A}''(z)(\mathrm{e}^{-Y} - 1)
    -4 \Sc v_2(z).
\end{equation}
Outside the boundary layer, the concentration decays to zero. Therefore, in
order for the concentration fields to match as
$Y\to\infty$, it must be the case that $4\Sc v_2 + \bar{A}'' \equiv 0$.
Note that, due to incompressibility, $v_2 = -\frac{1}{2} w_1'$. Consequently,
\begin{equation}
    \bar{A}(z) = \mathcal{I}(z) + \alpha z - \beta, ~~ \mathrm{where}~~
    \mathcal{I}(z) = 2 \Sc \int_0^z w_1(\tilde z) \,\mathrm{d}\tilde z,
    \label{eq:A and I}%
\end{equation}
with $\alpha$ and $\beta$ constants of integration.  The first of these is
necessarily zero, since the boundary conditions require $\bar{A}(z)$ to be an
$L_z$-periodic function. The second constant, $\beta$, is set by conservation of
concentration, which mandates that the integral of $\hat{c}$ over the whole
domain vanishes.  Therefore, $\beta = \langle \mathcal{I} \rangle$ 
and we are left with the following
asymptotic solution for the streamwise-averaged concentration field, which we
write in full, up to $O(1)$:
\begin{equation}
    \bar{c}(Y,z) 
    = 2\mathrm{e}^{-Y}\left[
    \frac{v_s}{\kappa}
    +
    \Sc\left( \int^z_0 w_1(\tilde z) \, \mathrm{d}\tilde z
    -\frac{1}{L_z}\int_0^{L_z}\!\! \int_0^z w_1(\tilde z)\,\mathrm{d}\tilde z \,\mathrm{d}z
    \right)
    \right].
    \label{eq:high vs asym}%
\end{equation}
Since the sign of $\bar{A}'(z)$ is everywhere identical to the sign of $w_1(z)$, the
maxima and minima of $\bar{A}(z)$ lie at cross-stream stagnation points $z = z^*$.  In
particular, maxima must occur when $w_1'(z^*) < 0$, meaning that sediment
preferentially accumulates at interfaces between rolls where spanwise velocity is
oriented towards $z^*$ from both sides, just as we have seen from the illustrative
plots of $EQ_1$ in figure~\ref{fig:cfields}.

\begin{figure}
    \includegraphics[width=\columnwidth]{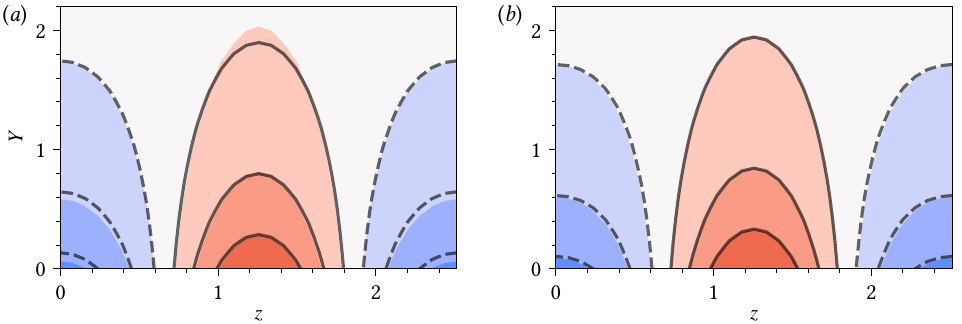}
    \caption{Concentration fields of $EQ_1$ in the boundary layer at the lower
    wall, for high $v_s/\kappa = $~(\emph{a})~$14$ and (\emph{b})~$40$.  Filled
    contours show the perturbation after taking a streamwise average,
    $\bar{c}(Y,z) - c_0(Y)$, with values spaced at intervals of $2$ and centred
    around zero (white).  Black contour lines at the same levels show the $O(1)$
    component of the asymptotic solution from Eq.~\eqref{eq:high vs asym}.
    Dashed and solid lines indicate negative and positive values respectively.
    }
    \label{fig:highvs_casym}
\end{figure}

The sediment distribution within the boundary layer can be visualised more
clearly by plotting contours of the streamwise-averaged perturbation to the base
concentration. In figure~\ref{fig:highvs_casym}, we plot these data for $EQ_1$ at
$v_s/\kappa =$~(\emph{a}) $14$ and (\emph{b}) $40$, obtaining excellent
agreement with the corresponding asymptotic solution of Eq.~\eqref{eq:high vs
asym}.

\subsection{Sediment transport}
So far we have seen that the velocity structures of ECS maintain a distribution
of sediment that differs from the laminar base profile in ways that are
analytically tractable in the regimes of low and high settling velocity. It is
natural to ask how this affects the transport of sediment in the channel. This
question depends on the reference frame of an observer monitoring the flow.  In
applications, it is typical for the bottom boundary to be stationary from an
observer's perspective, rather than the centreline $y=0$, as is the case for our
configuration.
The total streamwise flux $Q_x$, 
measured in a frame with stationary bottom boundary, is
\begin{equation}
    Q_x =
    \frac{1}{2}\int_{-1}^1 \langle \overline{(u + 1) c} \rangle \, \mathrm{d}y,
    \label{eq:uc}%
\end{equation}
which gives a bulk measure for the rate at which sediment is transported by the
flow. The portion, $\hat{Q}_x$, of this flux that is attributable to the presence of the ECS, is
\begin{equation}
    \hat{Q}_x = Q_x - \frac{1}{2}\int^{1}_{-1}(y+1)c_0\,\mathrm{d}y.
    \label{eq:Qxhat}%
\end{equation}
Another important value is the vertical flux $Q_y$, given by
\begin{equation}
    Q_y =
    \frac{1}{2}\int_{-1}^1 \langle \overline{vc} \rangle
    \,\mathrm{d}y,
    \label{eq:vc}%
\end{equation}
since this quantifies the amount of sediment uplifted by the ECS against the
flux from particle settling and thereby dictates the distribution of sediment
available for streamwise transport.

\begin{figure}
    \begin{center}
    \includegraphics{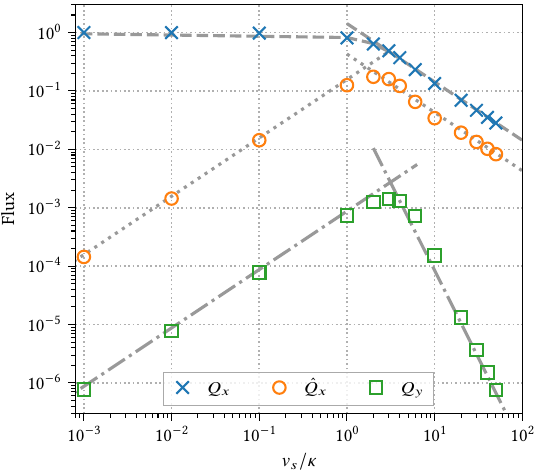}%
    \end{center}
    \caption{Sediment fluxes for $EQ_1$ at different $v_s/\kappa$ values
    (coloured points).
    The grey dashed lines are $Q_x = 1 - 0.176 v_s/\kappa$ and $Q_x = 1.429
    \kappa / v_s$,
    computed (and given to 3~s.f.) from the
    asymptotic approximations in Eqs.~\eqref{eq:uc low vs asym} 
    and~\eqref{eq:uc high vs asym} for
    streamwise fluxes in the limiting settling regimes. Their counterparts in dotted
    grey are the same lines shifted down by $1-v_s/(3\kappa)$ [see
    Eq.~\eqref{eq:Qxhat low vs}] and $\kappa / v_s$ respectively, 
    thereby giving the
    flux $\hat{Q}_x$ attributable to the ECS perturbation field. 
    The grey dash--dotted lines are $Q_y = 8.72\times 10^{-4} v_s/\kappa$ and $Q_y = 0.0841(\kappa/v_s)^3$ -- the asymptotic predictions for the vertical
    fluxes, given in Eqs.~\eqref{eq:vc low vs asym} and~\eqref{eq:vc high vs
    asym} respectively.
    }
    \label{fig:uc}
\end{figure}

In figure~\ref{fig:uc}, we show the fluxes for $EQ_1$ at different
values of settling velocity.
Total streamwise sediment flux [Eq.~\eqref{eq:uc}] is plotted with blue crosses.
This is a decreasing function of $v_s/\kappa$, because higher settling
velocities decrease the amount of sediment in the upper portions of the channel,
leading to lower rates of transport. Maximum flux occurs in the limit of
vanishing settling velocity, where the particle concentration becomes
homogeneous. With orange circles, we plot $\hat{Q}_x$ -- the corresponding
values of the streamwise sediment flux that may be attributed to the ECS
[Eq.~\eqref{eq:Qxhat}].  This component decays to zero in both extremal regimes
of settling velocity, attaining its maximum at $v_s/\kappa \approx 2$, as does
the vertical sediment flux [Eq.~\eqref{eq:vc}], plotted with green squares.  The
non-monotonicity of the fluxes $\hat{Q}_x$ and $Q_y$ is consistent with the
following physical interpretation. At low $v_s/\kappa$, the ECS can do little to
redistribute the homogeneous base concentration profile and consequently has a
negligible effect on the sediment transport. Increasing $v_s/\kappa$ towards
intermediate values enhances the impact of the ECS, whose vertical fluxes
maintain sediment concentrations in the channel that would otherwise settle
towards lower values.  However, at high settling velocities, accumulation of
sediment at the lower wall dominates and the trend reverses.
These insights generalise to other solutions and can be made precise by using
our asymptotic analyses of the passive concentration fields
in~\S\ref{sec:passive low vs} and~\S\ref{sec:high vs} to derive theoretical
estimates of the limiting behaviour in each regime.  These are the grey lines in
figure~\ref{fig:uc}, which are in excellent agreement with the data in both
regimes.  Their formulae are given below.

In the low settling velocity regime, the concentration field has the form $c = 1
+ (v_s/\kappa) \bar{c}_1 + \ldots$, with $\bar{c}_1$ set by
Eq.~\eqref{eq:lowvs final}. Therefore, to leading order in $v_s/\kappa$, the
vertical flux is
\begin{equation}
    Q_y = \frac{v_s}{\kappa} 
    \cdot
    \frac{1}{2}
    \!\int_{-1}^1 \langle \bar{v} \bar{c}_1 \rangle \,\mathrm{d}y.
    \label{eq:vc low vs asym}%
\end{equation}
The integrand in this expression is independent of $v_s/\kappa$.
Its value may be obtained for any ECS by numerically solving Eq.~\eqref{eq:lowvs final}
for $\bar{c}_1$ and using an appropriate quadrature to evaluate the integrals.
This was computed for $EQ_1$ in figure~\ref{fig:uc}, using the solution obtained
in the previous subsection [plotted in figure~\ref{fig:lowvs higher re}(\emph{a})].
Similarly, the streamwise flux may be written as
\begin{equation}
    Q_x = 1
    + 
    \frac{v_s}{\kappa}
    \cdot
    \frac{1}{2}
    \int_{-1}^{1}
    \langle
    -y\bar{u}+
    \bar{u}\bar{c}_1
    \rangle
    \,\mathrm{d}y. 
    \label{eq:uc low vs asym}%
\end{equation}
The first term in this expression, combined with the $\langle-y\bar{u}\rangle$ part of the
integral comprises the transport of the base
concentration field, which is positive.
The final term, corresponding to the transport of the concentration
perturbation, is expected to be negative, since the vortex structure of the ECS,
which advects low speed streaks together with high concentrations (and vice
versa, see figures~\ref{fig:eq1} and~\ref{fig:cfields}), causes
$\bar{u}$ and $\bar{c}_1$ to be inversely correlated.
However, the total transport due to the ECS, which is
\begin{equation}
    \hat{Q}_x = Q_x - 1 + \frac{1}{3}\frac{v_s}{\kappa} =
    \frac{v_s}{\kappa}\cdot\frac{1}{2}\int_{-1}^1 -y\langle \bar{\hat{u}} +
    \bar{u}\bar{c}_1 \rangle \,\mathrm{d}y,
    \label{eq:Qxhat low vs}%
\end{equation}
to leading order, is positive in figure~\ref{fig:uc} because the advection of
the base concentration by the streamwise velocity perturbation [first part of
the integrand in~\eqref{eq:Qxhat low vs}] is positive and outweighs the
contribution from the transport of the concentration perturbation.

In the limit of high settling velocity, the sediment localises towards a
boundary layer of thickness $\varepsilon = \kappa / v_s \ll 1$ at the lower wall. 
We determine the sediment fluxes to leading order in $\varepsilon$ by
substituting the expanded velocity fields of Eqs.~(\ref{eq:vwc
expansions}\emph{a--c}), together with the 
asymptotic solution of Eq.~\eqref{eq:high vs asym},
into the transport integrals and simplifying. Firstly, the streamwise flux
[Eq.~\eqref{eq:uc}] becomes
\begin{equation}
    Q_x = \frac{\kappa}{v_s} \left(
    1 + \langle u_1 \rangle
    \right)\!
    \label{eq:uc high vs asym}%
\end{equation}
The first term of this expression is the flux for purely laminar flow and the
second gives the particular contribution of the ECS (i.e.\ $\hat{Q}_x = Q_x -
\kappa / v_s$). For $EQ_1$, this latter
term comprises 30\% of the total flux, while for $EQ_2$ the proportion is
roughly 60\%. Using the same methods for the vertical flux, we obtain
\begin{equation}
    Q_y = \frac{2\Sc}{\Rey}\left(\frac{\kappa}{v_s}\right)^{\!3}
    \frac{1}{L_z}
    \int_0^{L_z}v_2\left(\mathcal{I} - 
    \langle \mathcal{I} \rangle
    \right)
    \mathrm{d}z,
    \label{eq:vc high vs asym}%
\end{equation}
where the $z$-dependent function $\mathcal{I}$ was defined in Eq.~\eqref{eq:A
and I}.  The rather acute inverse cubic dependence on settling velocity in this
case arises from the fact that within the $O(\kappa/v_s)$ boundary layer over
which the sediment is distributed,
the vertical velocity is $O(\kappa^2/v_s^2)$.

\section{Stratified regime ($\Ri_b > 0$)}
\label{sec:stratification}%
Since, due to settling, the concentration of particles is always higher at the
bottom of the channel and lower at the top, a stable stratification is induced
in the flow when the mass loading is nonzero. This couples the velocity
field to $c$ by energetically penalising vertical motions. We study the effect
that this has on ECS by using parametric continuation to trace out families
of solutions as $\Ri_b$ varies away from zero.  As seen in the previous
subsection, the concentration fields in the passive limit do not possess any
symmetries with respect to inversion of the wall-normal direction (except in the
limit $v_s/\kappa \to 0$, where $c\equiv 1$ throughout the domain) because sediment preferentially
accumulates at the bottom wall.  Therefore, the introduction of buoyancy effects means
that the upper and lower halves of the channel are no longer equivalent with
respect rotation around the $z$ axis. This breaks the `shift-and-rotate'
symmetry $\mathcal{R}$ of the ECS velocity fields
[Eq.~\eqref{eq:shift-and-rotate}], which otherwise forces
solutions to have vanishing bulk velocity. 
Consequently, as $\Ri_b$ deviates from
zero, states generally possess nonzero bulk velocity and become
become travelling waves with $a \neq 0$.

\subsection{Solutions with $v_s/\kappa = 1$}
\label{sec:vk 1}%
\begin{figure}
    \includegraphics[width=\columnwidth]{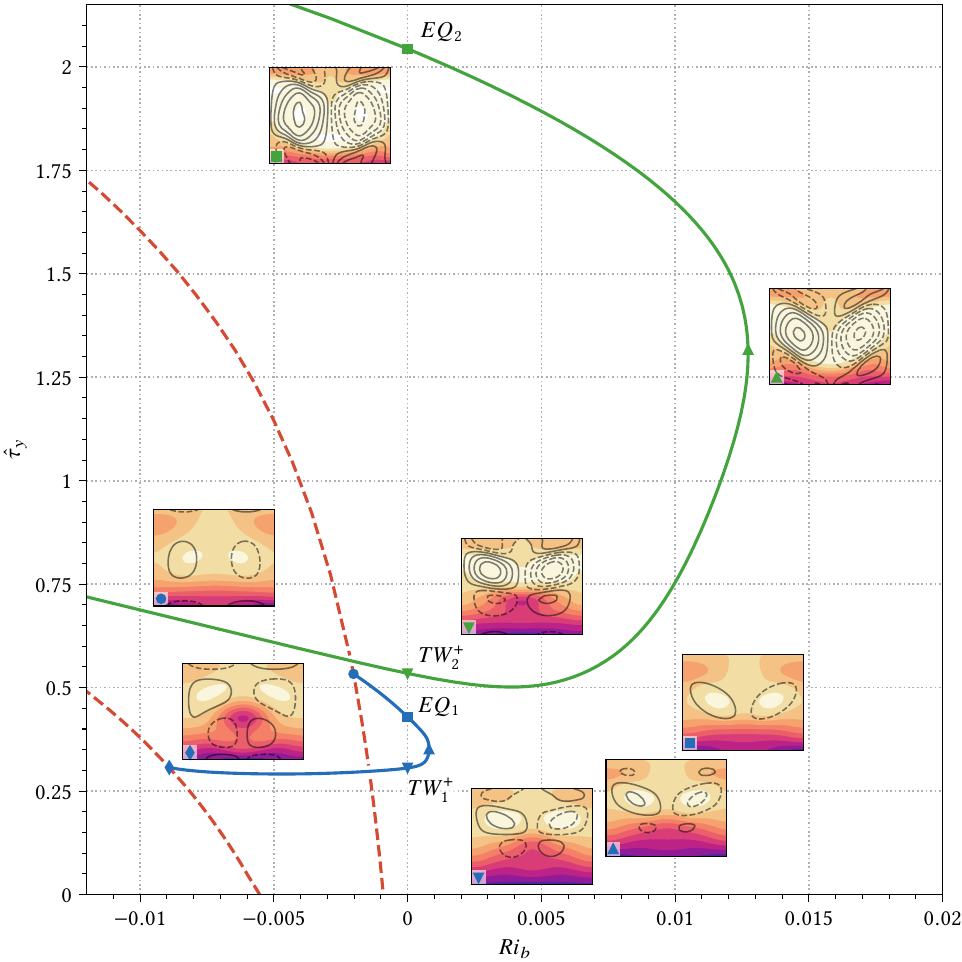}
    \caption{%
    Parametric continuation of $EQ_1$ (blue) and $EQ_2$ (green) in $\Ri_b$ as a
    function of mean excess wall-normal stress $\hat\tau_y$, for $v_s/\kappa = 1$.  The
    red dashed curves are two-dimensional roll solutions that bifurcate off from
    linear instabilities of the base flow at $\Ri_b = -9.08\times 10^{-4}$ and
    $-5.53\times 10^{-3}$.  Inlaid figures display data for states at the
    corresponding points indicated on the solid curves.  Filled contours show
    $\partial \bar c / \partial y$ using $10$ equispaced intervals from
    $-2.2$~(dark)
    to $0$~(light). These are overlaid with positive (solid) and negative
    (dashed) isolines of the mean streamwise
    vorticity $\partial \bar{w}/\partial y - \partial \bar{v}/\partial z$,
    at the values $\pm 0.08 n$ for $n = 1,\ldots,5$.
    The velocity fields of the states with $\Ri_b \geq 0$ are also
    plotted later, in figure~\ref{fig:bif megafig}.
    }%
    \label{fig:vk1 contRi}%
\end{figure}

Our numerical continuations begin away from either of the limiting regimes of low and
high settling velocity, by studying solutions with $v_s/\kappa = 1$, which we
take as an initial illustrative parameter choice. In figure~\ref{fig:vk1 contRi},
we plot some solution families connected to $EQ_1$ and $EQ_2$ at $v_s/\kappa =
1$ in terms of $\Ri_b$ versus the mean wall stress of the
perturbation to laminar flow $\hat{\tau}_y$.
%
%
%
Both $EQ_1$ (solid blue) and $EQ_2$ (solid green) emerge from saddle-node
bifurcations in $\Ri_b$, at different positive $\Ri_b$ values (upward-pointing
triangles).  However, in the particular projection shown in the plot, they are
not directly connected. This contrasts with continuation in $\Rey$ in
homogeneous plane Couette flow ($\Ri_b = 0$), where they are the lower and upper
branches of a saddle-node in $\Rey$~\citep[at $\Rey \approx 218.5$,
see][]{Gibson2009}.  As $\Ri_b$ increases, the stress of both states decreases
from the values of $EQ_1$ and $EQ_2$ respectively.  They reach their respective
saddle-nodes at $\Ri_b = 8.01\times 10^{-4}$ and $0.0127$, before turning back
towards lower $\Ri_b$.  It is noteworthy that these maximum $\Ri_b$ values are
relatively small, indicating that relatively weak buoyancy effects are
sufficient to disrupt unstable equilibria in this case.  Both solution families
have positive wave speed $a>0$ when they pass back through $\Ri_b=0$, giving
rise to two unstratified travelling waves, which we call $TW_1^+$ and $TW_2^+$.
These two states were originally discovered by~\cite{Viswanath2008}
and~\cite{Gibson2009} respectively, though not through direct connections to
$EQ_1/EQ_2$.  Our labelling convention for these waves diverges
from~\cite{Gibson2009} (who called them $TW_2$ and $TW_3$) in order to emphasise
their connection with the equilibria and distinguish them from symmetry-related
solutions with negative wave velocity ($TW_1^-$ and $TW_2^-$, discussed below),
since this becomes necessary when the states are continued in $\Ri_b$.

Inlaid plots in figure~\ref{fig:vk1 contRi} show the
mean streamwise vorticity (using solid and dashed contours for positive and
negative values respectively) and mean wall-normal concentration gradient
(filled contours) of the states at various points along the solution branches. 
In each case, concentration gradients are strongest at the lower wall, where
sediment preferentially accumulates, due to settling.
The two branches of each saddle-node indicate two ways in which states
adapt to the presence of stratification.
The trends are clearest along the $EQ_2/TW_2^+$ branches, but are evident on
both. From $EQ_2$ to the
corresponding saddle-node, the strength of the primary central vortex pair
decreases and the structure recedes away from highly stratified regions. 
Conversely, in the case of $TW_2^+$, which is localised towards the upper
channel, its vorticity increases towards the saddle-node and the concentration 
becomes more homogenised, especially in the mid-channel where the primary
vortices reside.

\begin{figure}
    \includegraphics[width=\columnwidth]{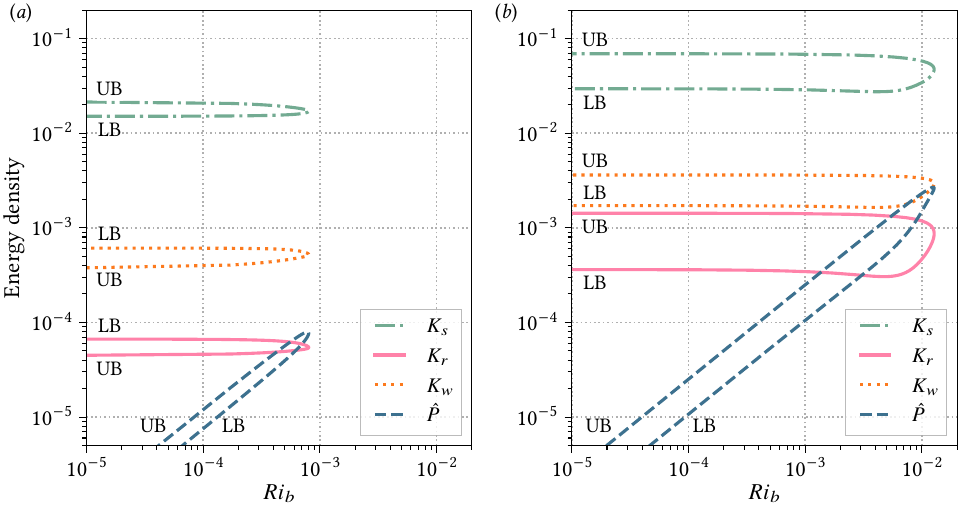}
    \caption{Energy density components associated with the perturbation fields
    of the (\emph{a})~$EQ_1/TW_1^+$ and (\emph{b})~$EQ_2/TW_2^+$ solution
    branches.  The labels `UB' and `LB' respectively indicate which parts of
    each curve correspond to the upper and lower branches in figure~\ref{fig:vk1
    contRi}.}
    \label{fig:vk1 contRi energy}%
\end{figure}
This picture can be complemented with a look at the energetics.  The total
energy density $E_{\textit{tot}}$ of each solution was given previously in
Eq.~\eqref{eq:energy density}. 
This can be decomposed into parts
associated with the streak, roll and wave components of the self-sustaining
process by substituting $\vect{u} = (\bar{u}-y,0,0)^T + (0,\bar{v},\bar{w})^T +
\vect{u}_w$, with $\vect{u}_w = \vect{u} - \bar{\vect{u}}$, into
Eq.~\eqref{eq:energy density} and separating out the terms corresponding to
these three pieces.
    This gives
    \begin{equation}
        E_{\textit{tot}} = 
        E_0
        +\frac{1}{2}\int_{-1}^{1}y\langle\overline{\hat{u}}\rangle\mathrm{d}y
        + K_s
        + K_r
        + K_w
        + \hat{P},
        \label{eq:Etot decomp}%
    \end{equation}
where $E_0 = 1/6 + \Ri_b(\kappa/v_s - \coth(v_s/\kappa))$ is
the total energy density of the base profile, $K_s$, $K_r$, $K_w$ are the
streak, roll and wave kinetic energy densities, and $\hat{P}$ is the potential
energy density of the concentration perturbation.
These constituents are defined by
\begin{subequations}
\begin{gather}
        K_s = \frac{1}{2}\int_{-1}^{1}\frac{1}{2}
        \big\langle\overline{\hat{u}^2}
        \big\rangle
        \,\mathrm{d}y,
        \quad
        K_r =
        \frac{1}{2}\int_{-1}^{1}\frac{1}{2}\langle
        \bar{v}^2 + \bar{w}^2
        \rangle\,\mathrm{d}y,\\
        K_w =
        \frac{1}{2}\int_{-1}^{1}
        \frac{1}{2}\langle\overline{\vect{u}_w\cdot\vect{u}_w}\rangle
        \,\mathrm{d}y,
       \quad
       \hat{P} =
        \frac{\Ri_b}{2}\int_{-1}^{1}
        \big\langle\overline{y\hat{c}}\big\rangle\,\mathrm{d}y
        .
\end{gather}
    \label{eq:ssp energies}%
\end{subequations}
The remaining term in Eq.~\eqref{eq:Etot decomp} (second on the right-hand side)
accounts for the net loss in kinetic energy due to the reorganisation of the
base shear profile by the ECS and is consequently negative.
The perturbation components
are plotted in figure~\ref{fig:vk1 contRi
energy}.
We see that the states are largely unaffected by the presence of density
gradients as $\Ri_b$ increases, until the potential energy of the perturbation
reaches the size of the roll kinetic energy.  Though this represents a
negligible portion of the total flow energy, only one of the three SSP
components needs to be fully disrupted to prohibit the existence of these
turbulence-supporting solutions.  In this case, the rolls, feel the
energetically penalising effects of stratification through their vertical
component, causing them to adapt in the ways already outlined above.  This
interaction ultimately sets the scale of the maximum $\Ri_b$ and mirrors
observations in linearly stratified shear flow~\citep{Eaves2015,Olvera2017}.  A
comparison of figures~\ref{fig:vk1 contRi energy}(\emph{a}) and~(\emph{b})
indicates that the $EQ_2$ branch withstands higher $\Ri_b$ values than $EQ_1$
because its rolls are an order-of-magnitude stronger.  Following some additional
continuations at $v_s/\kappa = 1$, where we explore connections with
previous plane Couette flow studies and
converge some new states, we shall return to address the question of the
maximum bulk Richardson number for other settling velocities in \S\ref{sec:vk <<
1 Rib > 0} and \S\ref{sec:vk >> 1 Rib > 0}.

Although $\Ri_b$ and $v_s$ are constrained to have the same signs for the
physical scenarios of this study, it is nevertheless informative to continue the
branches into the $\Ri_b < 0$ half-plane.  We find that $EQ_1$ is formed via a
bifurcation at $\Ri_b= -2.02\times 10^{-3}$ from a branch of streamwise
invariant roll solutions (dashed red curves in figure~\ref{fig:vk1
contRi}). This branch emerges from the first eigenmode of the laminar base flow
to turn unstable when $\Ri_b$ is decreased from zero -- a single
counter-rotating vortex pair -- which occurs at $\Ri_b = -9.08\times 10^{-4}$.
Likewise, $TW_1^+$ bifurcates off a streamwise-invariant branch of rolls (dashed
red) at $\Ri_b = -8.91\times 10^{-3}$, in this case two vortex pairs stacked
vertically in the channel, arising from an unstable mode at $\Ri_b = -5.53\times
10^{-3}$. The corresponding linear stability problem is covered in
Appendix~\ref{appendix:linear stability}. These observations are similar to
continuations of performed by \cite{Olvera2017} in stratified plane Couette
flow, who observed the $EQ_7$ state bifurcating from a set of sheared
Rayleigh-B\'enard convection rolls.  However, in our case, broken $\mathcal{R}$
symmetry implies that the 2D roll branches can carry nonzero bulk streamwise
velocity and the bifurcations off these branches break streamwise invariance to
create families of 3D travelling waves with nonzero $a$.  In contrast to $EQ_1$
and $TW_1^+$, the $EQ_2/TW_2^+$ family does not immediately appear to connect to
a nearby roll branch, at least as far as $\Ri_b = -0.055$ (whereupon we
terminated our continuations). 

\begin{figure}
    \includegraphics[width=\columnwidth]{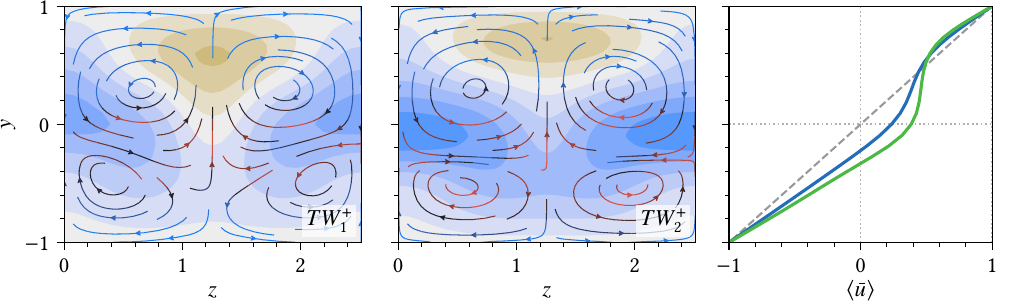}
    \caption{
        Travelling wave solutions $TW_1^+$ and $TW_2^+$ at $\Ri_b = 0$.  The
        first two panels show filled contours of the streamwise velocity
        perturbation $\bar{u} - y$, coloured from brown (negative) to blue
        (positive), using intervals of size $0.05$ centred around zero (white).
        Additionally,
        streamlines of $(\bar{v},\bar{w})$ are overlaid and coloured 
        according to $|(\bar{v},\bar{w})|$,
        from blue~(zero)
        to red. The red maximum value is $0.023$ for $TW_1^+$ and $0.049$ for
        $TW_2^+$~(2~s.f.).
        The rightmost panel
        plots the streamwise velocity after averaging in $x$ and $z$, 
        for $TW_1^+$ (solid blue) and $TW_2^+$ (solid green).  The
        dashed grey line is the streamwise velocity of laminar flow,
        i.e.\ $\langle \bar{u} \rangle = y$.
    }
    \label{fig:tws}%
\end{figure}

Figure~\ref{fig:tws} shows the travelling wave solutions $TW_1^+$ and
$TW_2^+$.
As already indicated by the corresponding inserts in figure~\ref{fig:vk1 contRi},
the solutions consist of two pairs of stacked vortices, with the stronger pair
of the two residing in the upper channel. These vortices redistribute momentum
by advecting faster positive streamwise velocity from the top of the channel
downwards and likewise bringing some velocity that is slower than the base flow
towards the top. Because this takes place preferentially in the upper channel,
the net effect is to skew the bulk velocity of the solutions positive.  This is
clear from the rightmost panel in figure~\ref{fig:tws}, which shows that $\langle
\bar{u} \rangle$ in both cases predominantly exceeds the corresponding value for
laminar flow. Furthermore, the two travelling wave velocities are positive, with
$a = 0.396$~($TW_1^+$) and $0.465$~($TW_2^+$). However, by the symmetry of the
equations at $\Ri_b=0$, the states defined by $TW_1^- = \mathcal{R} \cdot
TW_1^+$ and $TW_2^- = \mathcal{R} \cdot TW_2^+$ are also exact solutions, with
corresponding velocities $a = -0.396$ and $-0.465$.  Since their primary
vortices lie in the lower portion of the channel, where sediment gradients are
higher, these negative travelling waves respond differently to variations in
$\Ri_b$. 

\begin{figure}
    \includegraphics[width=\columnwidth]{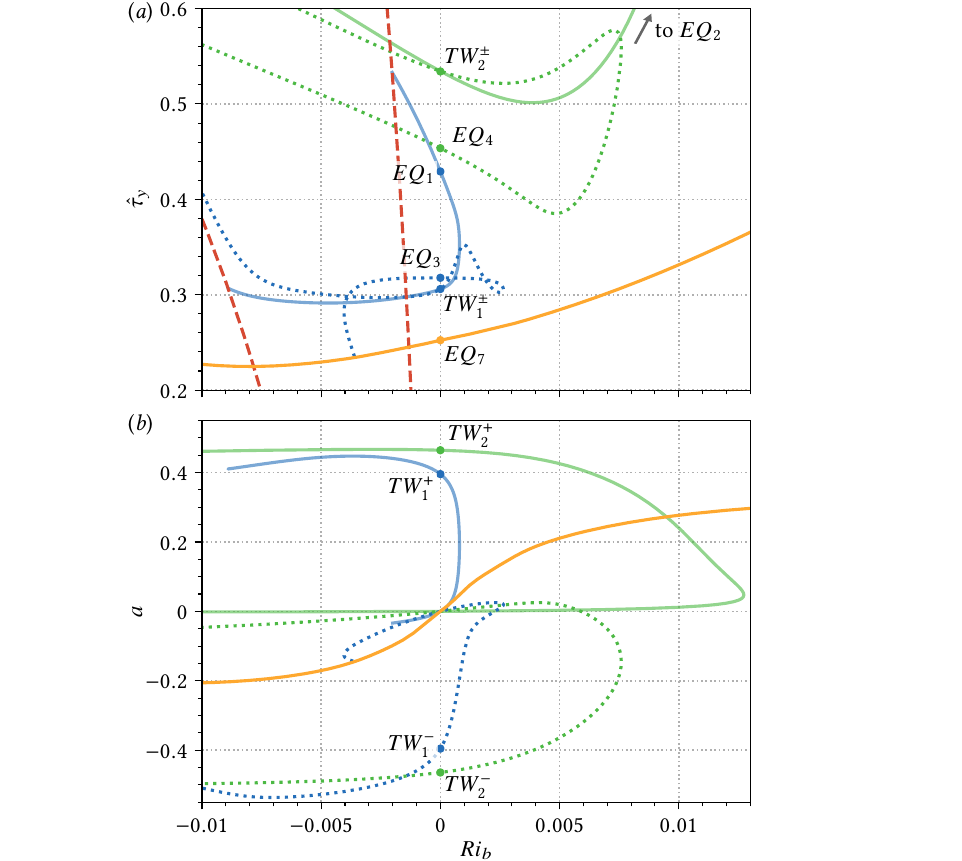}
    \caption{
    Parametric continuation of the $TW_1^-$ (dotted blue) and $TW_2^-$
    (dotted green) families and related states (solid lines).
    States are continued in $\Ri_b$ and plotted as a function of their
    (\emph{a})~mean wall-normal stress perturbation $\hat\tau_y$ and
    (\emph{b})~streamwise wave speed $a$.
    The dashed red curves in panel (\emph{a}) are 2D roll
    branches also shown in figure~\ref{fig:vk1 contRi}.
    In both plots, each labelled state is indicated with a filled circle.
}
    \label{fig:vk1 contRi extra waves}%
\end{figure}

Tracing out these related solution families uncovers a richer bifurcation
structure. 
In figure~\ref{fig:vk1 contRi extra waves}(\emph{a}), we plot the result of
continuing $TW_1^-$ and $TW_2^-$ in $\Ri_b$ (dotted curves), alongside the
$EQ_1$/$TW_1^+$ and $EQ_2$/$TW_2^+$ branches already shown in 
figure~\ref{fig:vk1 contRi}.
For reference, the corresponding streamwise wave speeds for each of the branches
are plotted directly below in figure~\ref{fig:vk1 contRi extra waves}(\emph{b}),
using the same horizontal axes.  This highlights that the symmetry of the
positive and negative travelling wave branches breaks as soon as $\Ri_b$
deviates from zero.  
For example, unlike the $TW_1^+$ and $TW_2^+$ branches, the $TW_1^-$ and $TW_2^-$ families
undergo a sign change of $a$ at positive $\Ri_b$, passing through equilibria at 
$\Ri_b = 2.3\times 10^{-3}$ and $\Ri_b = 5.9\times
10^{-3}$ (3~s.f.).
Both branches form a saddle-node in $\Ri_b$, attaining a maximum $\Ri_b$ and
connecting back to other equilibrium solutions at $\Ri_b = 0$ previously
identified by \cite{Gibson2009}.  Specifically, $TW_1^-$ connects to $EQ_3$ and
$TW_2^-$ connects to $EQ_4$.
Intriguingly, $EQ_3$ and $EQ_4$ are themselves connected, because they form a
saddle-node pair when unstratified plane Couette flow is continued in
$\Rey$~\citep{Gibson2009}, possibly hinting at a deeper relationship between this
pair and $EQ_1$/$EQ_2$.

When continued into negative $\Ri_b$, the $EQ_4/TW_2^-$ branches do not
obviously undergo successive bifurcations, at least within the range of our
computations ($\Ri_b>-0.055$).  The $TW_1^-$ branch appears to
asymptotically approach one of the 2D roll branches as $\Ri_b$ decreases,
though the branches remain disconnected. On continuing $EQ_3$,
we find that it connects to a family of solutions (solid orange) featuring
vortex pairs stacked vertically in the channel.
The new solution branch
possesses the additional half-cell shift symmetry
\begin{equation}
    \mathcal{T}:[u,v,w,c](x,y,z) \mapsto [u,v,w,c](x + L_x/2, y, z + L_z/2)
\end{equation}
which, when combined with $\mathcal{S}$ implies that states are mirror
symmetric in $z$.
The connection with $EQ_3$ occurs when this symmetry is broken in a pitchfork
bifurcation that gives rise to $EQ_3$ and its spanwise reflected counterpart.
Since spanwise reflections do not carry implications for the stratification, we
do not distinguish between such states.

\begin{figure}
    \includegraphics[width=\columnwidth]{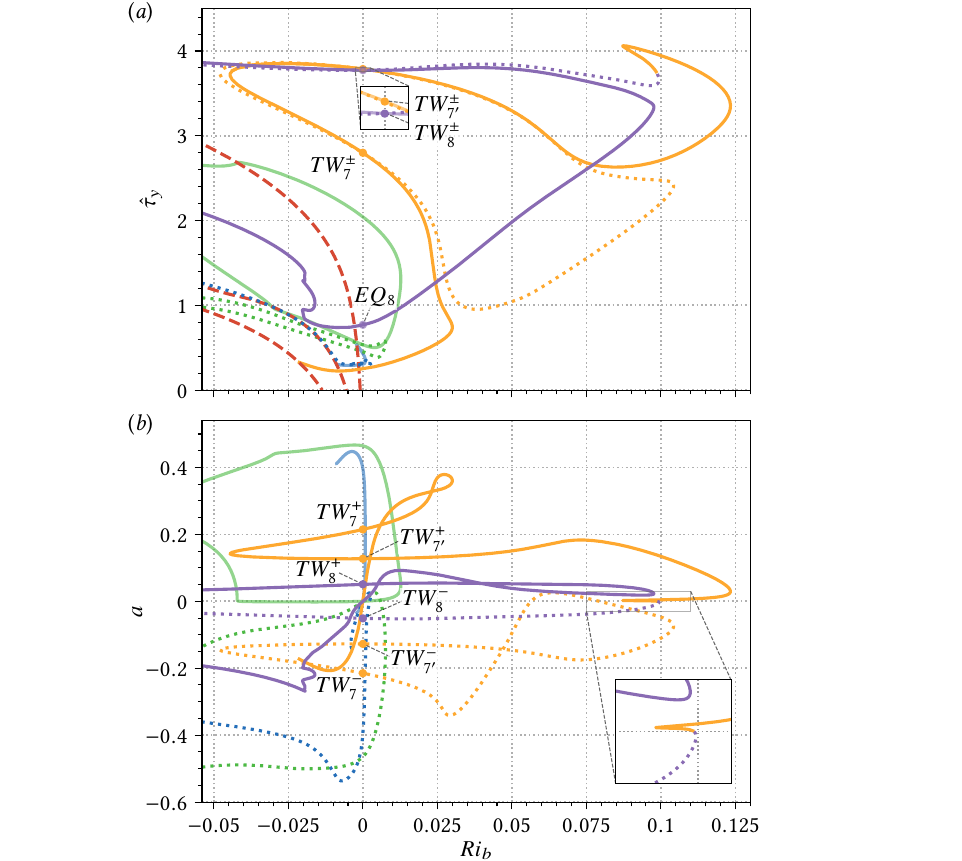}
    \caption{
    Bifurcation diagram for all solution branches identified at 
    $v_s/\kappa = 1$.
    Curves previously shown in figures~\ref{fig:vk1 contRi} and~\ref{fig:vk1
    contRi extra waves} are plotted together with
    the additional $TW_7^+$/$TW_{7'}^+$ (solid orange), $TW_7^-$/$TW_{7'}^-$
    (dotted orange),
    $EQ_8/TW_8^+$ (solid purple) and
    and $TW_8^-$ (dotted purple) families.
    States are continued in $\Ri_b$ and plotted as a function of their
    (\emph{a})~mean wall-normal stress perturbation $\hat\tau_y$ and
    (\emph{b})~streamwise wave speed~$a$.
    The dashed red curves are 2D roll
    solutions that bifurcate off from linear instabilities of the base flow.
    In addition to those identified in figure~\ref{fig:vk1 contRi},
    we include a branch that emerges at
    $\Ri_b = -0.0136$~(3~s.f.).
    In both plots, each labelled state is indicated with a filled circle.
    }%
    \label{fig:vk1 contRi full}%
\end{figure}

\begin{figure}
    \includegraphics[width=\columnwidth]{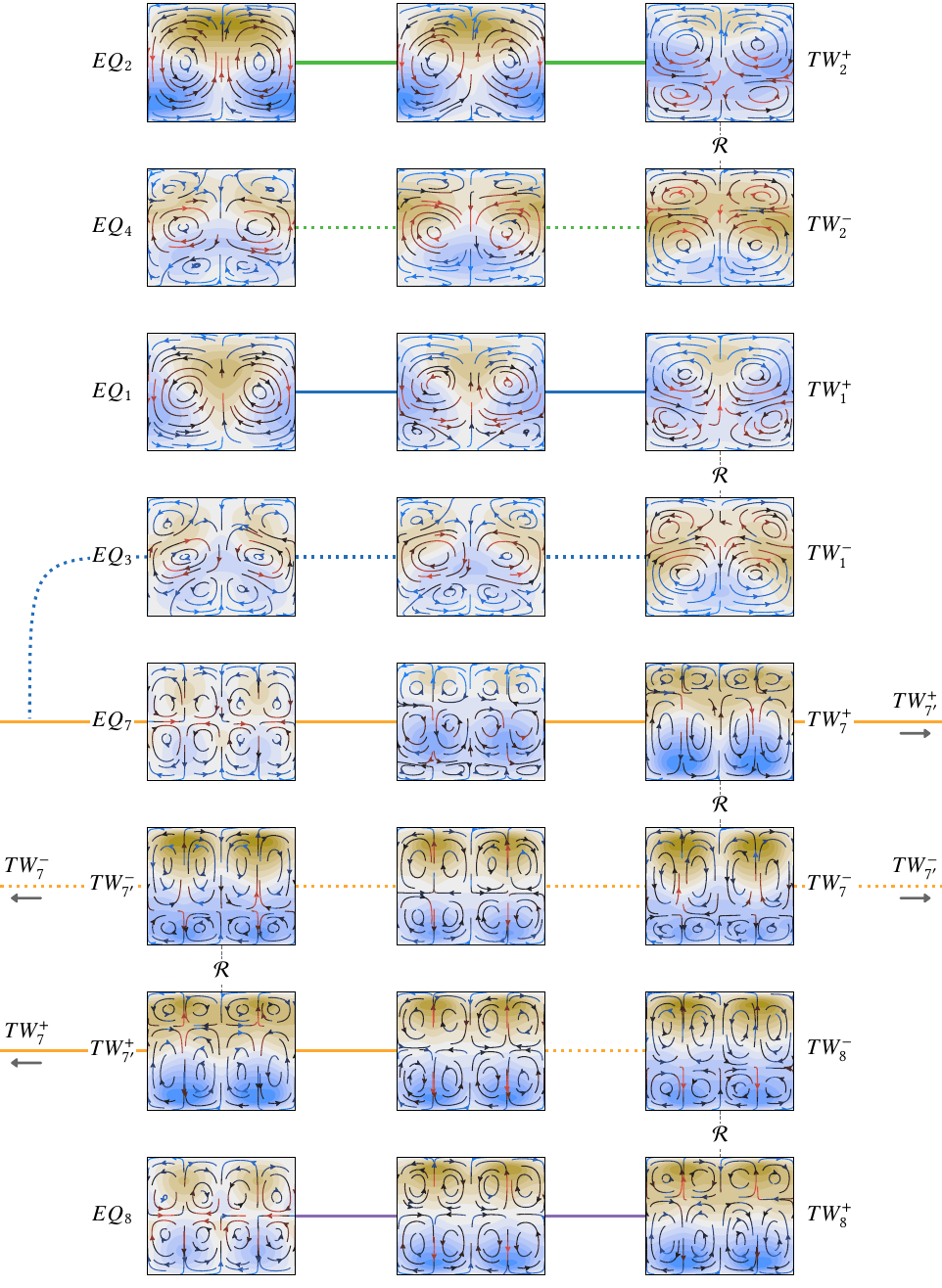}
    \caption{Diagram of relationships between states at $v_s/\kappa = 1$.  The
    first and third columns show named states at $\Ri_b=0$ that are connected by
    continuation in the $\Ri_b > 0$ half-plane. Connections are indicated by lines
    styled according to the corresponding curves plotted in figure~\ref{fig:vk1 contRi
    full}.
    Intermediate states plotted in the middle column are located at the
    maximum $\Ri_b$ attained along the relevant branch.
    Connections via $\Ri_b < 0$ are indicated in the margins.
    States related by the action of $\mathcal{R}$ are joined by a
    dashed grey line. 
    Each plot shows up to $15$ contours of $\bar{u} - y$ equispaced between
    $\pm0.75$ and streamlines of
    $(\bar{v},\bar{w})$, coloured in the same style as in figures~\ref{fig:eq1}
    and~\ref{fig:tws}.
    }%
    \label{fig:bif megafig}%
\end{figure}

The $\mathcal{T}$-symmetric branch leads to yet further equilibrium and travelling wave
solutions when continued in $\Ri_b$. We
plot the full bifurcation diagram obtained in figure~\ref{fig:vk1 contRi full} and
summarise the structures of all the solutions identified and the connections
between them in figure~\ref{fig:bif megafig}. 
Though complicated, we can systematically trace these families of exact
solutions and the exposition that follows is illustrated by both these
figures in concert.
The new branch emerges from a bifurcation off another 2D roll family at $\Ri_b =
-0.0215$~(3~s.f.) [the $(m,n)=(0,2)$ secondary mode, in the nomenclature of
Appendix~\ref{appendix:linear stability} linear stability analysis] and contains
the $EQ_7$ equilibrium -- a solution at $\Ri_b=0$ originally identified by
\cite{Gibson2009} and \cite{Itano2009}.  Following the branch further, it
reaches substantially higher $\Ri_b$ values than the solutions in figure~\ref{fig:vk1
contRi extra waves} and
weaves either side of $\Ri_b=0$ multiple times. The first two axis crossings give
rise to two new travelling wave solutions with $a>0$, which we call $TW_7^+$ and
$TW_{7'}^+$. Since these appear to be undiscovered homogeneous states for this
computational cell, we report their properties in Table~\ref{tab:tw properties}.
\begin{table}
  \begin{center}
\def~{\hphantom{0}}
  \begin{tabular}{lcccccccc}
      \vspace{0.1cm}
      & $||\cdot||$ & $E_{\textit{tot}}$ & $D$ & $\dim W^u$ & $\dim W_H^u$ & $a$
      & $\int_{-1}^{1}\langle \bar{u} \rangle\,\mathrm{d}y$ \\
      $TW_7^\pm$  & 0.3662 & 0.07547 & 3.798 & 32 & 3 & $\pm0.2148\phantom{0}$ & $\pm0.02198$ \\
      $TW_{7'}^\pm$  & 0.4027 & 0.06998 & 4.787 & 38 & 4 & $\pm0.1272\phantom{0}$ & $\mp0.01422$ \\
      $TW_8^\pm$  & 0.3957 & 0.07087 & 4.770 & 32 & 5 & $\pm 0.05104$ & $\mp0.02945$
  \end{tabular}
      \caption{Properties of the new travelling wave states when $\Ri_b = 0$.
      The table columns are as follows:
      $||\cdot||$ is the $L^2$-norm of the velocity perturbation $(\hat u, \hat
      v, \hat w)$; $E_{\textit{tot}}$ is the energy density~\eqref{eq:energy density}; $D$ is
      the dissipation~(\ref{eq:I and D}\emph{b}); $\dim W^u$ is the dimension of
      the unstable manifold; 
      $\dim W^u_H$ is the number of unstable eigenvectors that lie within the 
      symmetry group $H$ of the solution (in this case, the group
      generated by $\mathcal{S}$ and $\mathcal{T}$); $a$ is the wave speed and the final column reports
      the bulk velocity.
      }
  \label{tab:tw properties}
  \end{center}
\end{table}
They possess particularly high wall stress values and a greater number of unstable
eigenvalues than
any of the solutions presented by~\cite{Gibson2009}
or~\cite{Ahmed2020}, which may point to why they
have not been found in these prior studies.
Both the new states can be transformed, via
$\mathcal{R}$, to give their negatively-directed counterparts $TW_7^-$ and
$TW_{7'}^-$. Continuing these states reveals that they are
connected by an isola.

\begin{figure}
    \includegraphics[width=\columnwidth]{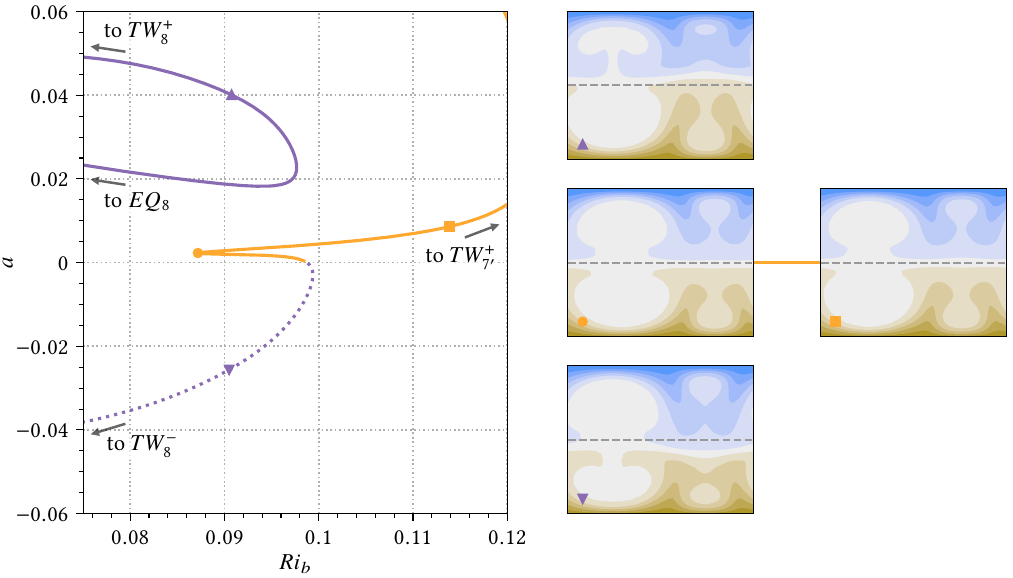}
    \caption{%
        Close-up view of the $TW_{7'}^+$/$TW_8^-$ and $TW_8^+$/$EQ_8$
        connections.
        The left-hand plot shows a portion of the continuation curves from
        figure~\ref{fig:vk1 contRi full}(\emph{b}), with the $TW_7^-$/$TW_{7'}^-$
        branches removed for clarity.
        The right-hand
        side panels show contours of $u(0, y, z)$ for the states at the locations
        indicated with the corresponding labels in the bifurcation diagram.
        For $\mathcal{R}$- and
        $\mathcal{T}$-symmetric states, such slices must be anti-symmetric with
        respect to reflection about the line $y=0$ (dashed grey).
        This property is approximately satisfied for the middle two panels,
        along the `central fork'
        and clearly broken for the `upper' and `lower' forks.
    }
    \label{fig:pseudo pitchfork}
\end{figure}

Following the $TW_7^+$/$TW_{7'}^+$ branch from $TW_{7'}^+$ towards higher $\Ri_b$,
it reaches a maximum at $\Ri_b = 0.123$~(3~s.f.), before turning and approaching
a structural feature suggestive of a pitchfork bifurcation at
$\Ri_b \approx 0.1$. 
This is highlighted in the inset panel of figure~\ref{fig:vk1 contRi full}.
Although the continuation curve is continuous through $a = 0$ here and does not
connect to other branches, we opt to
restyle the solution curve dotted purple where $a<0$.
As shown in figure~\ref{fig:pseudo pitchfork}, the yellow
`central fork' of the branch is approximately $\mathcal{R}$-symmetric as it
approaches $a=0$.
Since the branch is also $\mathcal{T}$-symmetric, this
implies that the solutions become close to being symmetric with respect to
rotation by $\pi$ around the $z$-axis, as highlighted in the
figure~\ref{fig:pseudo pitchfork} side panels.
(However, note that even at $a=0$, where the state becomes an equilibrium, this
symmetry cannot be exact, since $\mathcal{R}$ is not a symmetry of the
governing equations when both $v_s$ and $\Ri_b$ are nonzero.)  As the branch
continues to $a<0$, the symmetry becomes more obviously broken and this section
of the branch connects through to a new negatively-directed travelling wave
state, which we label $TW_8^-$ (see also figure~\ref{fig:vk1 contRi full}). 
The positively-directed $TW_8^+ = \mathcal{R}\cdot TW_8^-$ yields a separate
family that forms the other half of the disconnected pitchfork.  Bulk statistics
for this new solution pair are reported in Table~\ref{tab:tw properties}.  As
demonstrated by the contour plots in figure~\ref{fig:pseudo pitchfork}, the
`upper' and `lower' forks are approximately related by the $\mathcal{R}$
symmetry, lending credence to the idea that this structural feature is in
essence an $\mathcal{R}$-symmetry breaking pitchfork bifurcation, fractured into
disconnected pieces due to the presence of asymmetric stratification.  On
following the $TW_8^+$ branch further, we note that it ultimately connects to
$EQ_8$, which is the upper-branch counterpart of $EQ_7$ when these equilibria
are continued in $\Rey$~\citep{Gibson2009}.

\begin{figure}
    \includegraphics[width=\columnwidth]{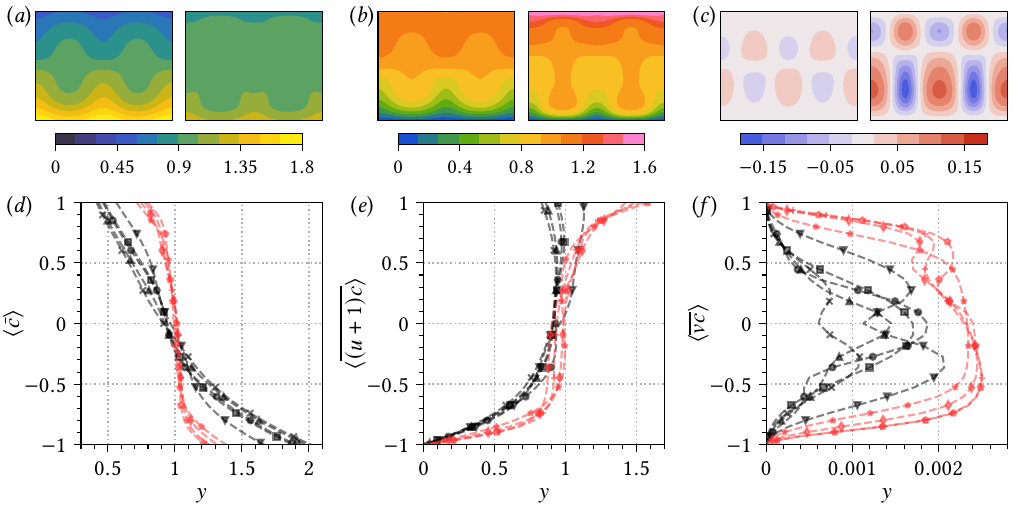}
    \caption{%
        Transport fluxes for low and high stress states with $\Ri_b=0$.
        (\emph{a}--\emph{c})~Contours of (\emph{a})~$\bar{c}$,
        (\emph{b})~$\overline{(u+1)c}$ and (\emph{c})~$\overline{vc}$ 
        for $EQ_8$ (left-hand panels) and $TW_8^+$ (right-hand panels).
        (\emph{d}--\emph{f})~Wall-normal profiles of
        (\emph{d})~$\langle\bar{c}\rangle$,
        (\emph{e})~$\langle\overline{(u+1)c}\rangle$ and
        (\emph{f})~$\langle\overline{vc}\rangle$,
        for the low stress states (dark grey dashed) $EQ_1$~(squares), $EQ_3$
        (upward-pointing triangles), $EQ_4$ (circles), $EQ_7$ (crosses), $EQ_8$
        (downward-pointing
        triangles) and high stress
        states (red dashed) $EQ_2$ (stars), $TW_7^\pm$ (diamonds), $TW_{7'}^\pm$
        (pluses), $TW_8^\pm$ (pentagons).
        }
    \label{fig:vk1 transport}
\end{figure}

The high wall-stress families emerging from $TW_7^\pm$, $TW_{7'}^\pm$ and
$TW_8^\pm$ have a noticeably different effect on the concentration field when
compared with the states with lower stress such as $EQ_1$ and $EQ_7$.
To demonstrate this, we compare states at $\Ri_b=0$
in figure~\ref{fig:vk1 transport}.
By way of illustration, panel~(\emph{a}) shows slices of $\bar{c}(y,z)$, for the
relatively low-stress $EQ_8$ state versus the high-stress travelling wave
$TW_8^+$.  In the latter case, the concentration field is relatively homogeneous
in the channel interior.  The corresponding implication for the sediment
transport is indicated in figures~\ref{fig:vk1 transport}(\emph{b,c}), which
respectively show mean streamwise and wall-normal fluxes for both states.
We see that the streamwise transport of $TW_8^+$ is higher throughout most of
the channel, facilitated by its stronger vortices, but it is far from
homogeneous -- fluxes are elevated near the high speed streaks where the
sediment is advected more rapidly.  In figures~\ref{fig:vk1
transport}(\emph{d--f}), we plot cross-stream averaged profiles for all of the
$\Ri_b=0$ solutions considered herein, dividing the data into low- and
high-stress groups. To the high-stress category, which includes our new
travelling wave solutions, we add $EQ_2$, since its excess stress is more than
double that of the other equilibria (see figure~\ref{fig:vk1 contRi full}).
(Other states from the study of~\cite{Gibson2009} that meet this criterion would
include $EQ_{11}$ and $EQ_{13}$.) The data are clearly separated in a manner
that mirrors our example contours given in figures~\ref{fig:vk1
transport}(\emph{a--c}), with the high-stress states exhibiting a more
homogeneous concentration field, and greater streamwise and wall-normal fluxes.
For small increments of $\Ri_b$ above zero these trends persist. Eventually
though, many of the lower stress states cease to exist, making it difficult to
compare the solutions collectively (figure~\ref{fig:vk1 contRi full}).  However,
the fact that each of the high-stress travelling wave branches can be continued
far higher $\Ri_b$ than the rest is notable and is perhaps due to the
homogenisation of the $c$ field, which lessens the effects of stratification in
the channel interior.

\subsection{Solutions with $v_s/\kappa < 1$}
\label{sec:vk << 1 Rib > 0}%
We now turn our attention to the effects of stratification on states away from
the illustrative case of $v_s/\kappa = 1$. To keep our numerical computations
manageable and to aid their reporting, we return to focusing on the
$EQ_1/TW_1^\pm$ and $EQ_2/TW_2^\pm$ families, with the anticipation that these
states are representative of the broader class of equilibria and travelling
waves underpinned by the SSP/VWI mechanism.

On decreasing the settling velocity, the concentration fields for each state 
are expected to become increasingly homogeneous, since they must
eventually become uniform in the limit $v_s/\kappa \to 0$, regardless of the
magnitude of $\Ri_b$. In \S\ref{sec:passive low vs}, we used this to determine that
in the passive regime ($\Ri_b=0$), the deviation of $c$ away from a homogeneous
state is $O(v_s/\kappa)$.  As $\Ri_b$ is increased away from this limiting case, we
can ask how large it must become in order for the effect of sediment-induced
stratification to be non-negligible.
Since
$\vect{u}$ and $c$ remain uncoupled to leading order until this regime is
reached, the magnitude of the density perturbation in the passive
case implies that the perturbation to the buoyancy force
is $O(\Ri_b v_s/\kappa)$, while $\Ri_b$ remains sufficiently small.
This reaches the size of the $O(\Rey^{-2})$ terms in the vertical momentum
equation when
$\Ri_b = O(\kappa v_s^{-1}\Rey^{-2})$. 
If $\Ri_b$ is increased beyond this regime, solutions that obey the VWI balances
must either cease to exist, or adapt to the presence of stratification by
modifying their structure to match the larger buoyancy forcing term. This can
happen by developing inner scales that enhance the spatial gradients of the flow
fields, i.e.\ localisation~\citep{Olvera2017}.

\begin{figure}
    \includegraphics[width=\columnwidth]{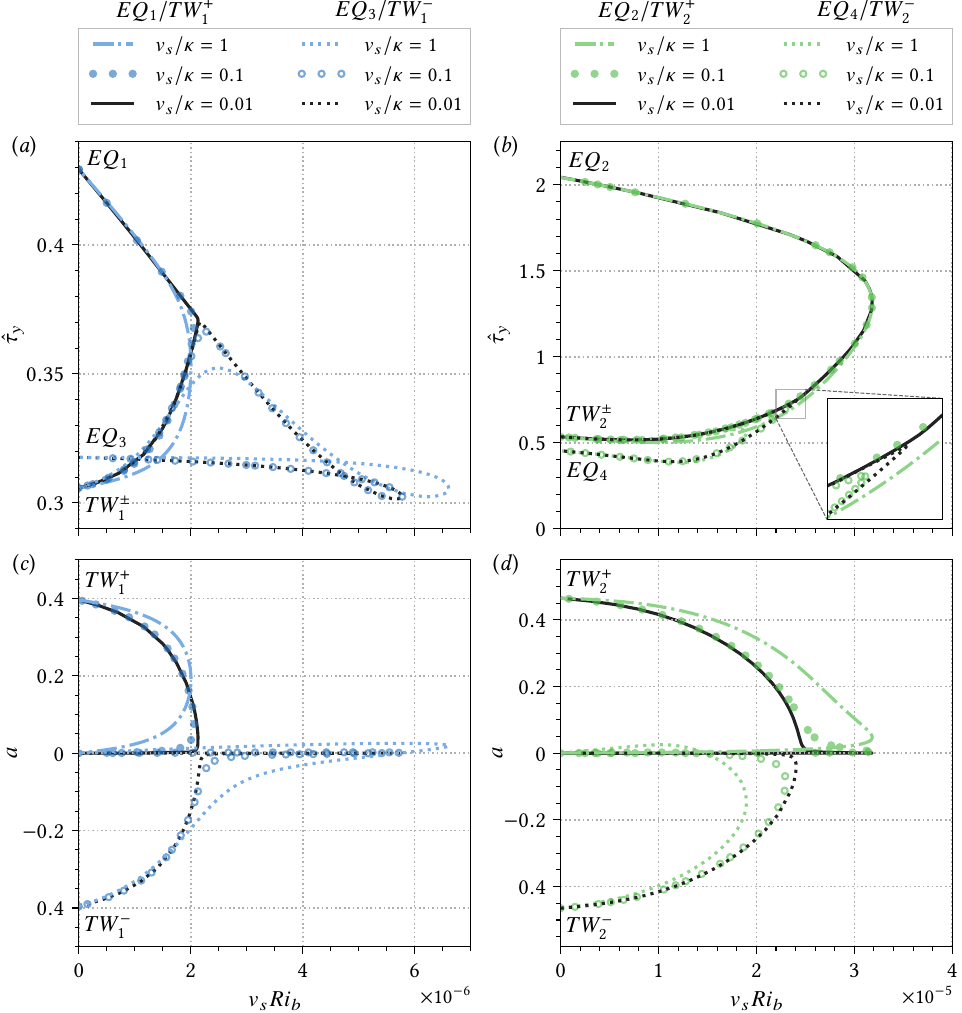}
    \caption{%
    Collapse of solution branches related to $EQ_1$ and $EQ_2$, as $v_s\to 0$.
    The two rows of plots show continuations in $\Ri_b$ versus
    the
    (\emph{a},\emph{b})~mean wall stress perturbation $\hat{\tau}_y$ 
    and (\emph{c},\emph{d})~streamwise wave
    velocity $a$,
    with the horizontal axes rescaled by $v_s$ to
    demonstrate collapse of the data.
    }
    \label{fig:lowvs}
\end{figure}

Within the minimal flow cell and moderate Reynolds number considered herein, 
this latter possibility does not occur.
Instead, the states exist up to a maximum $\Ri_b$ that is proportional to
$1/v_s$.
This is evidenced in figure~\ref{fig:lowvs}, in which continuations of our two example solution
families in $\Ri_b$ are plotted, for $v_s/\kappa$ decreasing from $1$ down to $0.01$.
We see that they collapse almost exactly onto a set of asymptotic curves when
rescaled under the combination $\Ri_b v_s$. Furthermore, the continuation
clarifies the relationship between the equilibria and travelling wave states.
The symmetry-related pairs $TW_1^\pm$ and $TW_2^\pm$ are created in subcritical
pitchfork bifurcations that are asymptotically realised as $v_s\to 0$, where the
system approaches invariance under $\mathcal{R}$. Since the equations decouple
in this limit, the $\Ri_b$ at which the bifurcation occurs is sent to infinity.  The equilibrium pairs
$EQ_1$/$EQ_3$ and $EQ_2$/$EQ_4$ are (asymptotically) connected either side of
their respective pitchforks, which may explain why the computations of
\cite{Gibson2009} showed that the portions of $EQ_3$ and $EQ_4$'s unstable
manifolds that lie within the solutions' symmetry groups are exactly one
dimension larger than those corresponding to $EQ_1$ and $EQ_2$ respectively.

\subsection{Solutions with $v_s/\kappa > 1$}
\label{sec:vk >> 1 Rib > 0}%
Increasing $v_s/\kappa$ beyond unity results in some structural changes to the
connections plotted in figures~\ref{fig:vk1 contRi} and~\ref{fig:lowvs}.
These are shown in figure~\ref{fig:medvstrend}.
\begin{figure}
    \includegraphics[width=\columnwidth]{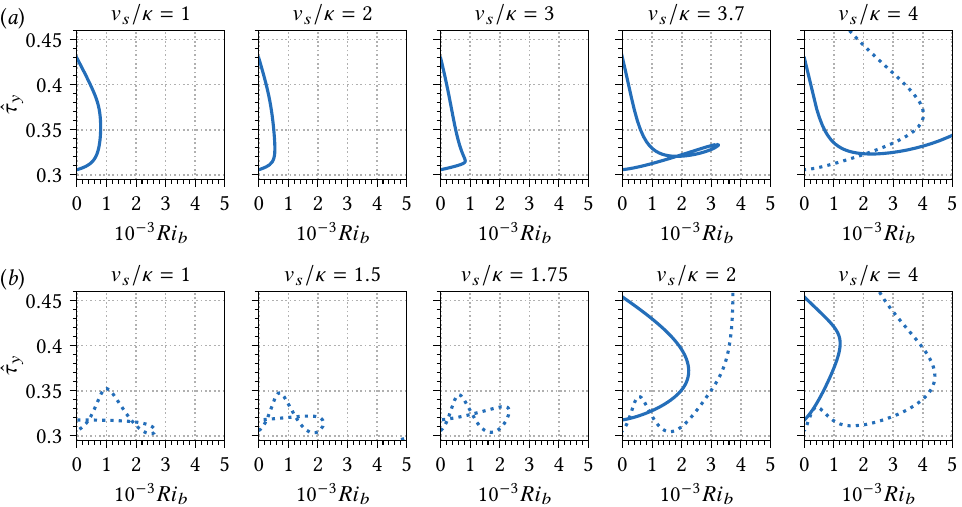}
    \caption{Sequence of continuations in $\Ri_b$ for increasing $v_s/\kappa$
    (as labelled).
    (\emph{a})~Continuation of $EQ_1$ and $TW_1^+$ (solid lines, first 4
    panels), which become disconnected by $v_s/\kappa = 4$ to leave separate
    branches. In the final panel, the $EQ_1$ branch is shown as a solid line and
    ultimately connects to $EQ_2$ (not shown) and the $TW_1^+$ branch
    is dotted and connects to $TW_2^+$ (not shown). 
    (\emph{b})~Continuation of $EQ_3$ and $TW_1^-$ (dotted lines, first 3 panels), which
    likewise become
    disconnected by $v_s/\kappa = 2$, with $EQ_3$ connecting to $EQ_4$ (solid
    lines) and
    $TW_1^-$ connecting to $TW_2^-$ (dotted lines).
    }
    \label{fig:medvstrend}
\end{figure}
Firstly, we see in row~(\emph{a}) that between $v_s/\kappa = 3.7$ and
$v_s/\kappa = 4$, the link between $EQ_1$ and $TW_1^+$ becomes severed.
At higher settling velocities, $EQ_1$ instead connects to $EQ_2$ (as it does
under $\Rey$ continuation) and $TW_1^+$ connects to $TW_2^+$.
Likewise, figure~\ref{fig:medvstrend}(\emph{b}), shows that between $v_s/\kappa
= 1.75$ and $2$, 
the $EQ_3/TW_1^-$ and $EQ_4/TW_2^-$ connections `swap partners', leaving
$EQ_3$/$EQ_4$ and $TW_1^-$/$TW_2^-$ as saddle-node pairs in $\Ri_b$ at higher
$v_s/\kappa$.

After these reorganisations at intermediate settling velocities, further
increases in $v_s/\kappa$ lead the solution branches towards a high settling
regime, in which states persist up to much higher $\Ri_b$ values.
Focusing on the $EQ_1/EQ_2$ branch, further continuations are shown
figure~\ref{fig:highvs}, up to $v_s/\kappa = 14$.
\begin{figure}
    \includegraphics[width=\columnwidth]{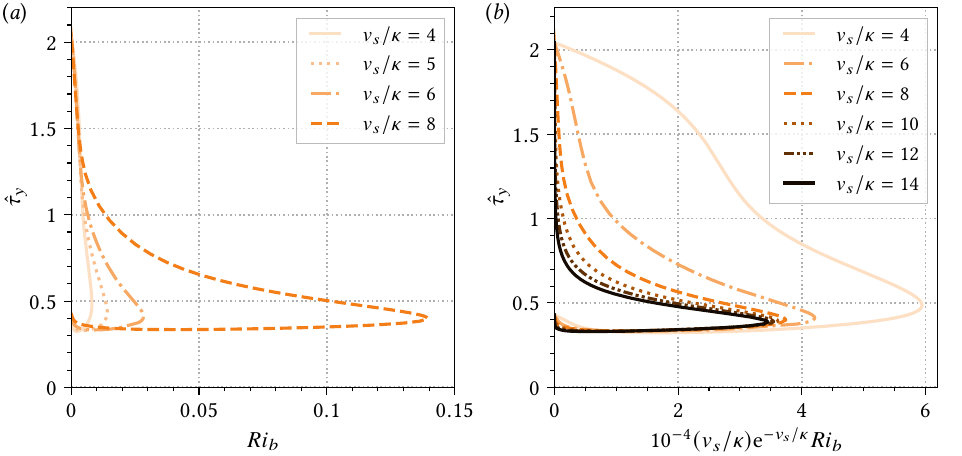}
    \caption{Continuation of the $EQ_1$/$EQ_2$ branch in $\Ri_b$, for
    high settling velocities.
    The solution curves 
    are coloured from light orange to black depending on $v_s/\kappa$.
    Panel (\emph{b})~demonstrates their collapse under rescaling $\Ri_b$
    by $(v_s/\kappa)\exp(-v_s/\kappa)$.
    }%
    \label{fig:highvs}
\end{figure}
Panel~(\emph{a}) of this figure demonstrates the trend, while panel~(\emph{b})
shows collapse of the curves under an exponential rescaling
that shall be rationalised later.  For example, denoting
the largest bulk Richardson number attained along the relevant solution branch
by $\Ri_{b,\textit{max}}$, there is an order-of-magnitude increase in this quantity
between the $v_s/\kappa = 6$ and $8$ cases, from $\Ri_{b,\textit{max}} = $ $0.028$ to
$0.14$ respectively.
Our prior analysis of the passive scalar regime points to why this might occur.
As observed in figure~\ref{fig:cfields}, increasing $v_s/\kappa$ from unity when
$\Ri_b = 0$ causes the sediment field to concentrate at the lower wall and
consequently dilutes
the remainder of the channel.  This suggests that for fixed
$\Ri_b\in(0,\Ri_{b,\textit{max}})$ and
increasing $v_s/\kappa$,
the coupled interaction between the concentration field and the velocity
structure must ebb away, as the buoyancy forces in the bulk of the channel
become weaker due to reduced sediment load.
This is demonstrated in figures and~\ref{fig:rifix vsup} and~\ref{fig:rifix
vsup conc}, in which lower branch states at $\Ri_b=0.002$ are plotted for
increasing $v_s/\kappa$.
\begin{figure}
    \includegraphics[width=\columnwidth]{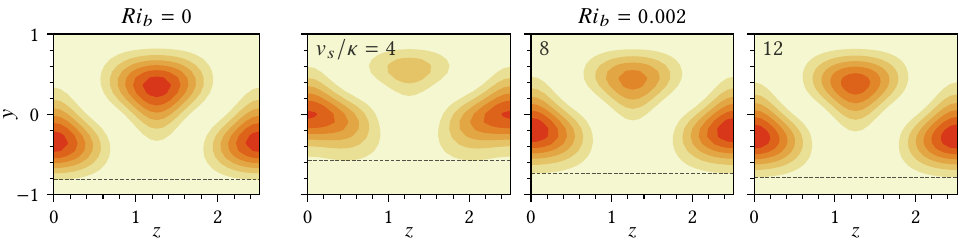}
    \caption{Effect of stratification as $v_s/\kappa$ is increased along the
    lower branch.  In each panel, seven filled contours of the
    streamwise-averaged perturbation
    kinetic energy $\frac{1}{2}(\overline{\hat{u}^2+\hat{v}^2+\hat{w}^2})$ are plotted,
    equispaced between~$0$ and $0.105$.  The leftmost panel shows $EQ_1$ at $\Ri_b = 0$
    and the trio of
    right-hand panels show the corresponding states obtained by continuing to
    $\Ri_b = 0.002$, with $v_s/\kappa = 4$, $8$ and $12$, as labelled.
    Grey dashed lines indicate the distance of the first contour from the lower
    wall in each panel.}%
    \label{fig:rifix vsup}
\end{figure}
\begin{figure}
    \includegraphics[width=\columnwidth]{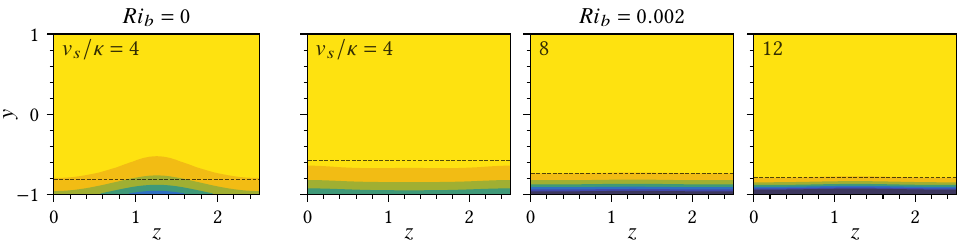}
    \caption{Streamwise-averaged concentration fields corresponding to the
    lower branch states in figure~\ref{fig:rifix vsup}. 
    Each panel shows seven filled contours of $\min(\overline{c}, 15)$.
    The applied thresholding removes very large values from the rightmost
    two plots below $y = -0.95$
    (which are not expected to affect the velocity fields significantly)
    so that the same contour scale can be used for each panel.
    The grey dashed lines lie at the same locations as those in figure~\ref{fig:rifix vsup}.}%
    \label{fig:rifix vsup conc}
\end{figure}
Relative to the passive case, which is also depicted in the two figures, the kinetic
energy of the velocity perturbation (figure~\ref{fig:rifix vsup}) is
preferentially located away from the highly concentrated boundary layers
(figure~\ref{fig:rifix vsup conc}). As $v_s/\kappa$ increases, the width of
these layers diminishes, reducing the effect of the sediment on the flow fields.

\begin{figure}
    \includegraphics[width=\columnwidth]{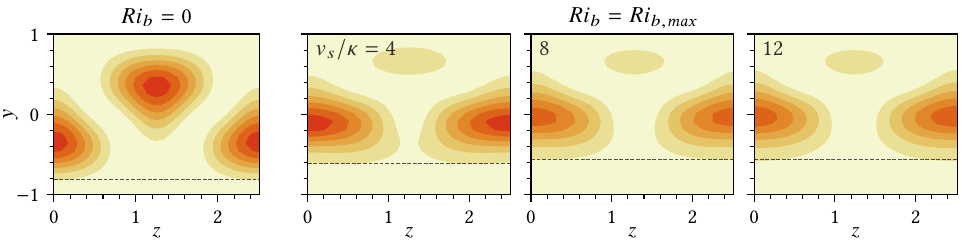}
    \caption{$EQ_1$/$EQ_2$ branch states at the saddle-node, where $\Ri_b =
    \Ri_{b,\textit{max}}$. Each panel shows
    streamwise-averaged perturbation kinetic energy contours,
    coloured in the same style as figure~\ref{fig:rifix vsup}, with the leftmost
    plot showing $EQ_1$ at $\Ri_b=0$ and the subsequent plots showing
    $v_s/\kappa = 4$, $8$ and $12$.
    Grey dashed lines indicate the distance of the first contour from the lower
    wall in each panel.
    }%
    \label{fig:highvs sn}
\end{figure}

Conversely, for fixed $v_s/\kappa$ and increasing $\Ri_b$, 
the vertical location for which an equivalent buoyancy effect is felt increases,
since this is proportional to $\Ri_b$. This enlarges the effective width of the
stratified lower region and ultimately drives states
away from the bottom wall.
At $\Ri_{b,\textit{max}}$, the velocity fields approach a limiting
solution as $v_s/\kappa$ becomes large.
This is summarised in figure~\ref{fig:highvs sn}, where we plot
the kinetic energy of the velocity perturbation at $\Ri_{b,\textit{max}}$ for
$v_s/\kappa = 4,8,12$, which is seen to be concentrated within the portion of
the channel above $y\approx -0.6$ for all three solutions.
Presumably, beyond this point the vortices are no longer able to adapt to the
presence of stratification by localising towards the upper part of the channel
and the solution branch must turn back on itself, as shown in
figure~\ref{fig:highvs}.

\begin{figure}
    \includegraphics[width=\columnwidth]{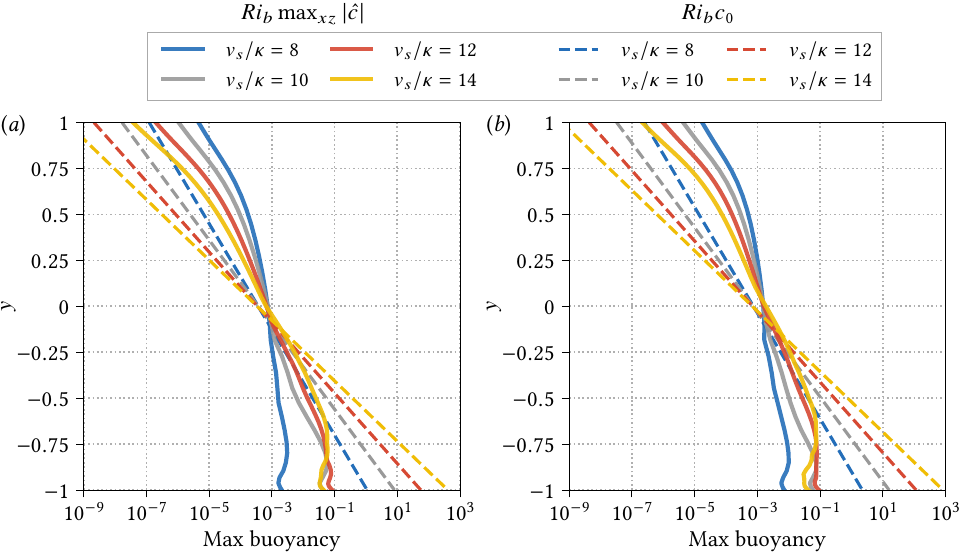}
    \caption{%
        Decay of the buoyancy term for high $v_s/\kappa$ states
        (\emph{a})~halfway along the $EQ_1$ 
        lower branch and (\emph{b})~at the saddle-node
        ($\Ri_b=\Ri_{b,\textit{max}}$).
        Dashed curves plot $\Ri_b c_0(y)$ for the corresponding cases.
    }
    \label{fig:strat cprofiles}
\end{figure}

Compared with the
low settling case (\S\ref{sec:vk << 1 Rib > 0}), the question of what the
asymptotic structures should be in the $v_s/\kappa\gg 1$ regime, is less clear cut.  On
increasing $\Ri_b$ from zero with $v_s/\kappa \gg 1$ fixed, the buoyancy force
becomes non-negligible when $\Ri_b\hat{c}$ is comparable to the dominant
balances in the vertical momentum equation.  
From our analysis of the passive case in \S\ref{sec:high vs}, this
occurs first near the bottom
wall, where $\hat{c}$ is $O(1)$ and the leading physics is controlled by
vertical diffusion of the $O(\Rey^{-1})$ velocity field. Therefore, the flow
momentum should begin to feel the presence of the sediment when $\Ri_b = O(\Rey^{-2})$.
However, we have seen that states cannot be fully disrupted by
near-wall stratification alone, so this cannot be the scale that
determines~$\Ri_{b,\textit{max}}$.
To identify the point beyond this scale, at which 
the velocity structure in the bulk of the channel becomes 
disrupted by stratification, we adopt an empirical approach.
Figure~\ref{fig:strat cprofiles}(\emph{a}) shows vertical profiles of the size of the
buoyancy perturbation (measured by $\Ri_b\max_{x,z}|\hat{c}|$) halfway along the
lower branch for the $v_s/\kappa =
8,10,12,14$ curves, alongside the
equivalent data for the contribution of the base concentration, $\Ri_b c_0$.
We see that, at least by $v_s/\kappa = 10$, the buoyancy perturbation curves
have a similar magnitude and dependence in the lower half of the channel and 
become considerably weaker in the upper half, as $v_s/\kappa$ increases.
The picture is qualitatively similar for states at $\Ri_{b,\textit{max}}$, whose
profiles are plotted in figure~\ref{fig:strat cprofiles}(\emph{b}) and possess
higher values overall, due to increased $\Ri_b$.
Based on these curves, we speculate that the asymptotic velocity fields drive a
concentration perturbation that collapses under an appropriate rescaling in the
lower half of the channel, but tends towards vanishing concentration in the
upper half as $v_s/\kappa\to\infty$. 
We note that, since the corresponding $c_0(y)$ curves all approximately intersect at a single point
$y=y^*$, the rescaling factor can be measured by leveraging the fact that $\Ri_b
c_0(y^*) \approx \mathrm{const}$ at this point.
This gives $\Ri_b \sim 1/c_0(y^*)\sim
\frac{\kappa}{v_s}\mathrm{e}^{\frac{v_s}{\kappa}(y^*+1)}$
as the scale that controls the structure of the high settling velocity states.
We estimate $y^*\approx 0$ from the figure.
As demonstrated in figure~\ref{fig:highvs}(\emph{b}), this is sufficient to collapse
the available data reasonably well.

\subsection{Summary}
\begin{figure}
    \includegraphics[width=\columnwidth]{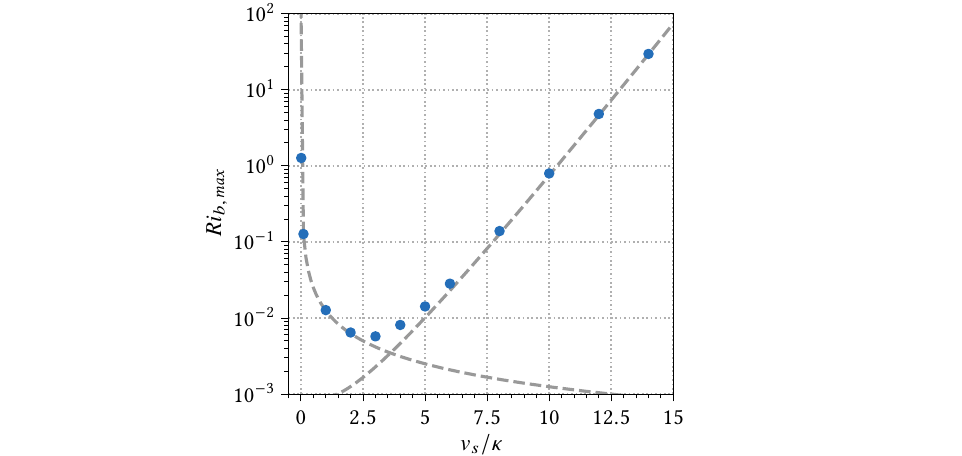}
    \caption{%
    Maximum bulk Richardson number attained by the $EQ_1$/$EQ_2$ family of
    states.  The fitted grey curves are $B_1 (v_s/\kappa)^{-1}$ and $B_2
    (v_s/\kappa)^{-1}\exp(v_s / \kappa)$ with $B_1 = 0.0125$ and 
    $B_2 = 3.4\times 10^{-4}$.
    }
    \label{fig:Ribmax}
\end{figure}

Taken together, our investigations of the $EQ_1/EQ_2$ family in both extremal regimes of settling
velocity imply that $\Ri_{b,\textit{max}}$ is non-monotonic with respect to
$v_s/\kappa$, because $\Ri_{b,\textit{max}}$ becomes unboundedly large as either
$v_s/\kappa \to 0$ or $v_s/\kappa \to \infty$.
This echoes our findings in the passive scalar regime, where it was seen that
either limit causes the concentration field to homogenise,
leading to non-monotonic sediment transport curves.
We summarise our continuation data in the stratified regime, by
plotting $\Ri_{b,\textit{max}}$ for each computed $v_s/\kappa$ in
figure~\ref{fig:Ribmax}.
Both limits are seen to agree well with the asymptotic scalings deduced
in~\S\ref{sec:vk << 1 Rib > 0} and \S\ref{sec:vk >> 1 Rib > 0} (grey curves).
Other states that are underpinned by the SSP/VWI mechanism in the passive scalar
regime can reasonably expect to scale similarly, at least in the low settling
regime. In the high settling regime, where geometric considerations appear to
play a role in constraining the interaction between the sediment and velocity
fields, it may be the case that solutions with different vortex morphologies,
such as $EQ_7$, behave quantitatively differently.  Nevertheless, the
scaling regimes demonstrated carry implications for the range of bulk Richardson
numbers that support turbulent flow and may be compared with prior DNS
explorations of the laminar--turbulent boundary, which are quantitatively
similar to figure~\ref{fig:Ribmax}~\citep{Cantero2012a,Langham2024}. This is
discussed in greater detail in the following section. We note additionally 
that the minimum of the $\Ri_{b,\textit{max}}$ curve
occurs at $v_s/\kappa \approx 2.5$, which is close to point at which vertical
and streamwise transport of the concentration perturbation is maximised for the
passive case (figure~\ref{fig:uc}).

\section{Discussion}
\label{sec:discussion}%
To recapitulate, we have computed and analysed equilibrium and travelling wave
solutions in plane Couette flow with a dilute suspension of sediment.  To keep
this initial investigation manageable we fixed $(\Rey,\Sc) = (400, 1)$ (save
briefly, in figure~\ref{fig:lowvs higher re}) and focused on the effect of
particle settling velocity for both passive and stratified sediment
concentration fields.

Low settling velocities ($v_s/\kappa \ll 1$) promote a nearly well-mixed
suspension. In this regime, 
the laminar base solution approaches a symmetric
linear concentration profile of $O(v_s/\kappa)$ gradient as $v_s/\kappa$
decreases.
Consequently, the flow physics is similar to linearly stratified plane
Couette flow. 
For passive sediment ($\Ri_b=0$),
the ECS concentration field is an
$O(v_s/\kappa)$ correction to the base profile, which
leads to vertical and downstream sediment fluxes of the
same order.  
When $\Ri_b$ is increased from zero, the size of buoyancy is then $O(\Ri_b
v_s/\kappa)$.
As a result, states are found to become more resilient to the effects
of stratification 
as $v_s/\kappa$ decreases, with the underlying vortex--wave
interaction becoming disrupted when $\Ri_b = O(\kappa v_s^{-1} \Rey^{-2})$.
This may be compared to the case of ECS with a linear base density
stratification imposed by Dirichlet boundaries
and $\Sc \ll 1$, for which the density perturbation
shrinks to $O(\Sc)$ and $\Ri_b= O(\Sc^{-1}\Rey^{-2})$ sets the
scale at which buoyancy modifies and eventually suppresses
states~\citep{Langham2020}. 
Note also, that while the effect of $\Sc$ has not been explicitly considered
herein, the linear analysis 
around \eqref{eq:marginal boundary vk << 1} in Appendix~\ref{appendix:linear
stability}  points toward the possibility that $\Ri_b=O(\kappa v_s^{-1}
\Sc^{-1}\Rey^{-2})$ is a more generally correct scale, at least for $\Sc$ not
too far from unity.

Even when settling is weak, vertical symmetry
breaking is important, with our continuations revealing that the $EQ_1$ and
$EQ_2$ states (Nagata's lower/upper branch equilibria) form the central prongs
of underlying pitchfork bifurcations that are asymptotically realised as
$v_s/\kappa \to 0$ and lead to travelling wave solutions of equal and opposite
velocities. 
This forms part of a broader picture of travelling wave pairs emerging
generically from equilibria that have their shift-and-rotate
$\mathcal{R}$ symmetry broken when $\Ri_b$ deviates from zero 
because their streamwise phase
is no longer fixed. Similar observations appear in the study of~\cite{Azimi2021}, who
considered homogeneous plane Couette flow with a
wall-normal suction term that also breaks $\mathcal{R}$ symmetry.
For intermediate settling velocity ($v_s/\kappa = 1$), we
catalogue various travelling waves that are obtained as a result of $\Ri_b$
continuation and symmetry transformations. Our thorough, yet inexhaustive
bifurcation diagrams, chart connections via travelling waves between known
equilibria and to unstable eigenmodes of the base flow at $\Ri_b < 0$, as well
as uncovering a new set of high wall stress travelling waves states that can be
continued back to $\Ri_b=0$.

Conversely, high settling velocities ($v_s/\kappa \gg 1$) result in the
formation of a sediment rich boundary layer at the bottom wall, with the
concentration decaying exponentially towards the upper channel.  For passive
sediment, the amplitude at the lower wall is given by the $O(v_s/\kappa)$ base
concentration profile, with an $O(1)$ correction driven by the streamwise rolls.
This leads to reduced sediment transport, with streamwise and wall-normal fluxes
that decay like $O(\kappa/v_s)$ and $O((\kappa/v_s)^3)$ respectively.  Under
stratification, states persist by withdrawing and sitting atop the highly
concentrated boundary layer. Consequently, they are able to attain very high
bulk Richardson numbers by remaining localised to the exponentially diluted
upper channel, with $\Ri_{b,\textit{max}}$ scaling
like~$\frac{\kappa}{v_s}\exp(v_s/\kappa)$.

Though they are usually performed at higher Reynolds number, previous direct
numerical simulations~(DNS) of particle-laden channel flows have pointed to
similar dependences on the settling velocity.  Most directly relevant are the
plane Couette DNS data of~\cite{Langham2024} at $(\Rey,\Sc)=(3125,1)$, in which
flows were seen to separate into regimes of stratified, but spatiotemporally
homogeneous turbulence (with essentially well-mixed sediment) at low settling
velocities, and a highly intermittent sedimentary boundary layer underneath
dilute unstratified turbulence, at high settling velocities.  Moreover, at
sufficiently high $\Ri_b$, these regimes are separated by regions of parameter
space that do not support turbulence, delineated by laminar--turbulent
boundaries of the form $\Ri_b \propto \kappa/v_s$ (for low $v_s$) and $\Ri_b
\propto \exp(0.65 v_s/\kappa)$ (for high $v_s$). The first of these
relations~\citep[which is also observed in gravity-driven channel flows,
see][]{Cantero2012a,Shringarpure2012}, matches the scale at which ECS cease to
exist at low settling velocities, supporting the view that these solutions are
needed for turbulent dynamics. The second relation differs from the bounds for
$v_s/\kappa \gg 1$ on the ECS identified herein only up to the empirically
determined part of the exponent ($0.65$).  Since this case appears to be
contingent on a wall-normal length that sets the minimum vortex size, it seems
probable that the Reynolds number and domain size play a role in determining
this latter scaling.  Simulations of pressure and gravity-driven channels
exhibit similar dependence of the concentration field on settling velocity and
can partially or fully laminarise when a boundary layer of sediment forms at the
bottom
wall~\citep{Cantero2009a,Cantero2009b,Cantero2012a,Shringarpure2012,Dutta2014}.

Looking ahead, it will be interesting to see whether the ECS paradigm can offer
new insights into sediment transport, since a deeper understanding of structures
within turbulence is needed to improve existing empirical models.  Our study has
demonstrated that focusing and advection of sediment by the streaks and rolls of
the ECS provide a mechanism for uplift and maintenance in suspension.  Though we
have left the investigation of dynamics for future study, the present analysis
suggests that the contribution of individual structures to bulk sediment fluxes
in an unsteady flow could be determined and theoretically rationalised.  It
seems plausible that, just as some ECS play a more important dynamical role than
others~\citep{Skufca2006,Gibson2008,Gibson2009,Park2015,Crowley2022}, particular
states may be more important for maintaining the suspension.  We have already
seen some evidence of this in the new high wall stress travelling waves
converged in \S\ref{sec:vk 1}, which feature elevated concentration levels in
the upper channel and correspondingly higher fluxes (see figure~\ref{fig:vk1
transport}).  Bursting periodic orbit
solutions~\citep{Kawahara2001,Viswanath2007} would also be promising to study in
this regard.  This could help to understand better the relative importance of
violent sediment ejections in suspension dynamics.

~\\
{\normalfont\fontfamily{LinuxLibertineT-TLF}\selectfont\large\fontseries{sb}\fontshape{n}\fontsize{10.5}{12}\selectfont
Acknowledgements.}
This research was financially supported by an EPSRC New
Horizons grant (EP/V049054/1), as well as Royal Society funding (APX/R1/180148)
and the computational facilities of the
Advanced Computing Research Centre, University of Bristol.

~\\
{\normalfont\fontfamily{LinuxLibertineT-TLF}\selectfont\large\fontseries{sb}\fontshape{n}\fontsize{10.5}{12}\selectfont Declaration of
interests.}
The authors report no conflict of interest.

\appendix%

\section{Energetic balance}%
\label{appendix:energy}%
In the dimensionless units for our system, the energy density of the mixture is
the sum of kinetic and gravitational potential contributions:
$E = |\vect{u}|^2/2 + \Ri_b y c$. The evolution of the first term
is obtained by taking the dot product of Eq.~\eqref{eq:governing 1} with
$\vect{u}$ and using incompressibility, to leave
\begin{equation}
    \frac{\partial~}{\partial t}\left(
    \frac{1}{2}|\vect{u}|^2
    \right) + \nabla \cdot \left(\frac{1}{2}\vect{u}|\vect{u}|^2\right)
    = -\nabla\cdot(p\vect{u})
    + 
    \nabla\cdot(\vect{\tau}\vect{u})
    -\vect{\tau} : \nabla\vect{u}
    - \Ri_b vc,
    \label{eq:ke}%
\end{equation}
where $\vect{\tau} = \Rey^{-1}[\nabla\vect{u} + (\nabla\vect{u})^T]$ is the deviatoric part of the
fluid stress tensor. Furthermore, Eq.~\eqref{eq:governing 3} implies that
\begin{equation}
    \frac{\partial~}{\partial t}(yc) + \nabla\cdot(yc\vect{u}) - vc
    = v_s\frac{\partial~}{\partial y}(yc) - v_s c
    + \kappa \left[
        \nabla\cdot(y\nabla c) - \frac{\partial c}{\partial y}
        \right].
    \label{eq:gpe}%
\end{equation}
Combining Eqs.~\eqref{eq:ke} and~\eqref{eq:gpe} thus gives an equation for the
energy density
\begin{equation}
    \frac{\partial E}{\partial t}
    +
    \vect{u}\cdot \nabla E
    = 
    \nabla\cdot(\vect{\tau}\vect{u})
    + \Ri_b \left[
        v_s \frac{\partial~}{\partial y}(yc) + \kappa\nabla\cdot(y\nabla c)
        \right]
    -\vect{\tau}:\nabla\vect{u}
    -\Ri_b\left(
    v_s c + \kappa \frac{\partial c}{\partial y}
    \right)
    - \nabla\cdot(p\vect{u}).
\end{equation}
The left-hand side of this equation is zero for an equilibrium state, or for a
travelling wave following the Galilean transformation $x \mapsto x - at$,
$u\mapsto u-a$.
Integrating the right-hand side over the whole domain therefore gives the
balance of total energy for a given state.
After simplifying using the boundary conditions, the parts that survive are
\begin{equation}
    \frac{1}{2\Rey}\left(\frac{\mathrm{d} \langle \bar{u} \rangle}{\mathrm{d}
    y}\bigg|_{y=1} + 
    \frac{\mathrm{d} \langle \bar{u} \rangle}{\mathrm{d} y}\bigg|_{y=-1}\right)
    =
    \frac{1}{2\Rey}\int_{-1}^1\langle\overline{\nabla\vect{u}:\nabla\vect{u}}\rangle\,\mathrm{d}y
    + \frac{\Ri_b}{2}\!\!\int_{-1}^1\left(
    v_s \langle \bar{c}\rangle + \kappa \frac{\mathrm{d} \langle \bar{c}
    \rangle}{\mathrm{d} y}
    \right)\,\mathrm{d}y.
\end{equation}
The final integral quantifies the gravitational loading of the sediment.  By
leveraging expressions obtained by averaging Eqs.~\eqref{eq:governing 1} and
\eqref{eq:governing 3} over the $x$ and $z$ directions, the
balance~\eqref{eq:energy balance} presented in the main text may be obtained.

\section{Linear stability}%
\label{appendix:linear stability}%
In this Appendix we compute linear stability modes of the laminar base flow.
As will be shown below, linear instability arises when $\Ri_b < 0$.
Such a parameter value is unphysical, given our
assumed sign convention $v_s>0$ [see Eq.~\eqref{eq:settling symmetry}], since it
would imply that the settling particles are positively buoyant.  However, as
noted in~\S\ref{sec:stratification}, some states at $\Ri_b \geq 0$ can be traced
via parameter continuation to bifurcations off the base flow at $\Ri_b < 0$.
Therefore, linear analysis is helpful for understanding the origin of these
states and may also prove useful for finding new ones.

We retain the convention of using hatted variables to denote perturbations to
the laminar base flow. If these are sufficiently small (i.e.\ $|\hat{\vect{u}}|,
|\hat{p}|, |\hat{c}| \ll 1$), then to leading order the governing
equations~\eqref{eq:governing 1}--\eqref{eq:governing 3} are linear, and become
\begin{subequations}
\begin{gather}
    \frac{\partial \hat{\vect{u}}}{\partial t} + y\frac{\partial
    \hat{\vect{u}}}{\partial x} + \hat v \vect{e}_x = -\nabla \hat p
    + \frac{1}{\Rey}\nabla^2 \hat{\vect{u}} - \Ri_b \hat c \vect{e}_y,\label{eq:lin prob 1}\\
    \nabla \cdot \hat{\vect{u}} = 0,\label{eq:lin prob 2}\\
    \frac{\partial \hat c}{\partial t} + y\frac{\partial \hat c}{\partial x}
    + \hat v \frac{\partial c_0}{\partial y}
    - v_s \frac{\partial \hat c}{\partial y} = \kappa \nabla^2 \hat c.\label{eq:lin prob 3}
\end{gather}
\end{subequations}
The pressure perturbation may be eliminated from the vertical momentum equation
by taking the divergence of Eq.~\eqref{eq:lin prob 1}
and substituting the resulting expression for $\nabla^2 \hat p$ into the
Laplacian of the vertical component of Eq.~\eqref{eq:lin prob 1}. Under
incompressibility [Eq.~\eqref{eq:lin prob 2}], this
simplifies to 
\begin{equation}
    \left(\frac{\partial~}{\partial t} + y \frac{\partial~}{\partial x}\right)
    \nabla^2 \hat v
    = \frac{1}{\Rey}\nabla^4 \hat v
    - \Ri_b\left( \frac{\partial^2 ~}{\partial x^2}
    + \frac{\partial^2 ~}{\partial z^2}
    \right)\hat c.
    \label{eq:lin prob v}%
\end{equation}
Equations.~\eqref{eq:lin prob 3} and~\eqref{eq:lin prob v} form an independent
coupled system for $\hat v$ and $\hat c$, upon which the remaining fields $\hat
u$, $\hat w$ and $\hat p$ depend.  Therefore, only $\hat v$ and $\hat
c$ are explicitly considered henceforth.
We decompose them into modes, via the Laplace and Fourier transforms, as the real
part of
\begin{subequations}
\begin{equation}
    \hat v(x,y,z) = \widetilde v(y)\mathrm{e}^{\sigma t + 2\pi\im(m x / L_x
    + n z / L_z)}, \quad
    \hat c(x,y,z) = \widetilde c(y)\mathrm{e}^{\sigma t + 2\pi\im(m x / L_x
    + n z / L_z)},
    \tag{\theequation\emph{a,b}}%
\end{equation}
\end{subequations}
where $\sigma$ is an arbitrary complex growth rate, $m$, $n$ are
positive integer values 
and $\widetilde v$, $\widetilde c$ are complex functions in the
wall-normal direction, which must obey the fundamental 
boundary conditions $\widetilde{v} = 0$
and $\kappa \partial \widetilde{c} / \partial y + v_s \widetilde{c} = 0$ at $y =
\pm 1$, as well as $\partial \widetilde{v}/\partial y = 0$ at $y = \pm 1$, which
follows from no-slip and the eliminated incompressibility equation.
Using these expressions, Eqs.~\eqref{eq:lin prob 3} and~\eqref{eq:lin prob v}
are converted into the following system of ordinary differential equations
\begin{subequations}
\begin{gather}
    \left[
        \frac{1}{\Rey} L^2 - (\sigma + \im \alpha y)
    \right] L^2 \widetilde{v} = -\Ri_b (\alpha^2 + \beta^2)
\widetilde{c},\label{eq:ode eigproblem 1}\\
    \left[
        \kappa L^2 - (\sigma + \im \alpha y) + v_s
        \frac{\mathrm{d}~}{\mathrm{d}y}
    \right]\widetilde{c} = \widetilde{v}\frac{\mathrm{d}c_0}{\mathrm{d}y},
\label{eq:ode eigproblem 2}%
\end{gather}
\end{subequations}
where $\alpha \equiv 2\pi m / L_x$, $\beta \equiv 2\pi n / L_z$ are the
perturbation wavenumbers and
$L^2 \equiv \mathrm{d}^2/\mathrm{d}y^2 - \alpha^2 - \beta^2$ is the transformed
Laplacian operator.

\begin{figure}
    \includegraphics[width=\columnwidth]{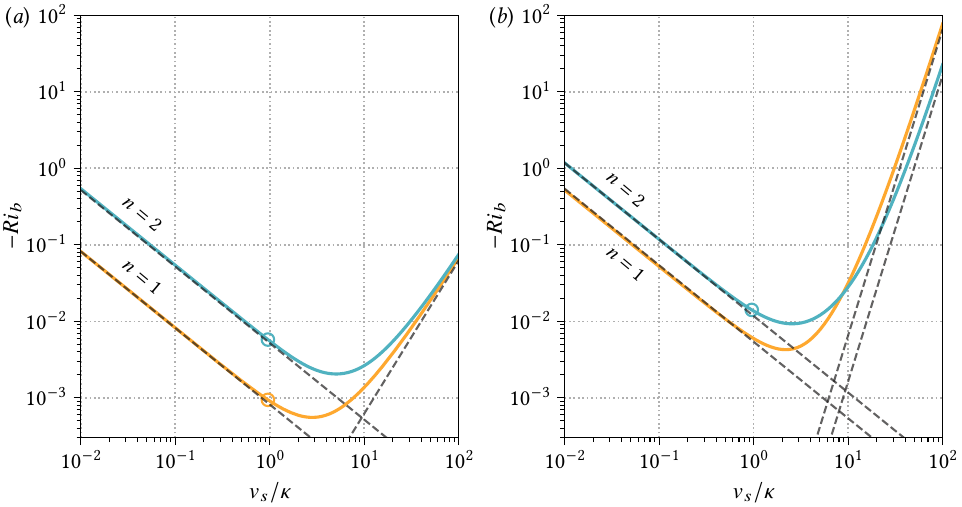}%
    \caption{Marginal stability curves for the (\emph{a})~primary and
    (\emph{b})~secondary streamwise-invariant modes with $n =
    1$ (orange) and $n = 2$ (blue).  The locations of the three modes at
    $v_s/\kappa=1$ that give rise to nonlinear solution families plotted in
    Fig.~\ref{fig:vk1 contRi full}(\emph{a}) are indicated with hollow circles.
    Also plotted for each family of modes are the theoretical curves derived in
    the limits of low (dash--dotted) and high (dashed) settling velocity.
    }%
    \label{fig:lin stab}%
\end{figure}

Given any wavenumber pair $(\alpha, \beta)$, this system may be solved numerically as an
eigenvalue problem for $\sigma$ and the corresponding perturbation modes
$\widetilde{v}$, $\widetilde{c}$. In this Appendix, we are particularly
concerned with the streamwise invariant case, $m = 0$, since these modes arise
from the computations in the main text.
The simplest such modes occur when $n = 1$, since it may be verified that when
$m=n=0$, Eqs.~\eqref{eq:ode eigproblem 1} and~\eqref{eq:ode eigproblem 2} admit
only the trivial solution $\widetilde{v}=\widetilde{c}=0$.  In each nontrivial
case, there is a discrete spectrum of wall-normal solutions to Eqs~\eqref{eq:ode
eigproblem 1} and~\eqref{eq:ode eigproblem 2}.  These can be ordered according
to their growth rate $\Real(\sigma)$. We refer to those with highest
$\Real(\sigma)$ as `primary' modes, second highest as `secondary', and so on.
The base flow is linearly stable when for all $m, n$, the primary mode has
$\Real(\sigma) \leq 0$.  For $\Rey = 400$ and $\Sc = 1/(\Rey\kappa)=1$, we
numerically the trace marginal stability boundaries, $\sigma = 0$, as a function
of $\Ri_b$ and $v_s$, for the primary and secondary modes with $(m,n) = (0, 1)$
and $(0,2)$, using the Chebfun software package~\citep{Driscoll2014}.
These curves are presented in Fig.~\ref{fig:lin stab}.
Unsurprisingly, each mode requires $\Ri_b < 0$ (i.e.\ unstable bulk
stratification), to become unstable.  Nevertheless, these curves provide a
useful check against our computations in the main text.  The loss of linear
stability of the different modes at $v_s/\kappa = 1$ leads to the bifurcations
presented in Figs.~\ref{fig:vk1 contRi} and~\ref{fig:vk1 contRi full}, where the
red dashed curves cross the horizontal axes. These points agree to at least $3$
s.f.\ with the corresponding locations on the marginal stability curves,
highlighted in Fig.~\ref{fig:lin stab}.

Furthermore, it is interesting to note that the linear stability problem
exhibits the same non-monotonic dependence on $v_s/\kappa$ as various properties
of the nonlinear solutions discussed in the main text.  Specifically, the flow
becomes more stable in both the asymptotic limits of low and high settling
velocity.  This is because in both these limits, the laminar base concentration
field becomes homogenised in the channel interior and the system approaches
plane Couette flow without particles, which is linearly stable.  As in the main
text, we can inspect both these regimes analytically.

Firstly, in the low settling velocity regime, $v_s/\kappa \ll 1$ and $c_0 = 1 -
(v_s/\kappa)y + \ldots$ to leading order. Substituting this into
Eqs.~\eqref{eq:ode eigproblem 1} and~\eqref{eq:ode eigproblem 2}, eliminating
$\widetilde{c}$, and considering only
the case of marginally stable streamwise invariant modes ($\sigma = \alpha = 0$)
gives
\begin{equation}
    L^6 \widetilde{v} = 
    \frac{v_s}{\kappa}\left(
    \Ri_b \Rey^2 \Sc \beta^2 - \frac{\mathrm{d}~}{\mathrm{d}y} L^4
    \right)\widetilde{v},
\end{equation}
where $L^4 = L^2L^2$ and $L^6 = L^2L^4$.
While the second term on the right-hand side could (in principle) be important,
we know that ultimately, $\Ri_b \Rey^2 \Sc \beta^2 \gg 1$ in this regime.
On neglecting this term, we see that the system parameters for each marginally
stable mode
are constrained to be the eigenvalues of the operator $\mathcal{L} \equiv
(\mathrm{d}^2/\mathrm{d}y^2 - \beta^2)^3/\beta^2$. That is,
\begin{equation}
    \mathcal{L} \widetilde{v} = \Lambda \widetilde{v}, \quad
    \mathrm{with}
    ~~
    \widetilde{v} = \frac{\mathrm{d}\widetilde{v}}{\mathrm{d}y} = 
    \frac{\mathrm{d}^5\widetilde{v}}{\mathrm{d}y^5} - 
    2 \beta^2 \frac{\mathrm{d}^3\widetilde{v}}{\mathrm{d}y^3}
    = 0
    ~~\mathrm{at}~~y=\pm 1,
    \label{eq:Lv = lamv}%
\end{equation}
where $\Lambda = (v_s/\kappa)\Ri_b\Rey^2\Sc$ and the final boundary condition is
determined by eliminating $\widetilde{c}$ from the no-flux condition on
$\widetilde{c}$. For each $\beta$, or equivalently each $n = 1,2,\ldots$, there
is a discrete spectrum of eigenvalues for this problem, which we index as
$\Lambda_{j,n}$ for $j= 1,2,\ldots$ and order in terms of increasing
magnitude (so that $j=1,2$ for the primary and secondary modes respectively).
Then the marginal stability boundary for each 
mode in the $v_s/\kappa \ll 1$ regime is given by the relation
\begin{equation}
    \Ri_b = \frac{\kappa}{v_s}
    \frac{\Lambda_{j,n}}{\Rey^2\Sc}.
    \label{eq:marginal boundary vk << 1}%
\end{equation}
Equation~\eqref{eq:Lv = lamv} is an ordinary differential equation with constant
coefficients, whose eigenvalues may be readily computed.
We find $\Lambda_{j,1} = -131, -830, \ldots$ and $\Lambda_{j,2}
= -865, -1879, \ldots$, and include the resulting asymptotic lines on
Fig.~\ref{fig:lin stab}, which demonstrate close agreement with the separately
computed stability modes.

In the case of high settling velocity, $v_s/\kappa \gg 1$, the sediment
concentration predominates within a boundary layer of characteristic thickness
$\varepsilon = \kappa / v_s$.
Since our interest is in the first few modes, we assume for the remainder that $\beta\ll
\varepsilon^{-1}$.
To leading order in $\varepsilon$, the base concentration field is $c_0(y) =
2\varepsilon^{-1}\exp[-(1+y)/\varepsilon]$.
The dominant balances for Eqs.~\eqref{eq:ode eigproblem 1} and~\eqref{eq:ode
eigproblem 2} are different inside and outside the boundary layer in this regime.
Therefore,
we develop separate asymptotic expansions and match them together.
In both cases, it is useful to rescale the concentration modes by defining
\begin{equation}
    \widetilde\phi(y) = \varepsilon \kappa
    \mathrm{e}^{(1+y)/\varepsilon}\widetilde{c}(y).
\end{equation}
Then, for $\sigma = \alpha = 0$, Eqs.~\eqref{eq:ode eigproblem 1}
and~\eqref{eq:ode eigproblem 2} become
\begin{subequations}
\begin{gather}
    L^4 \widetilde{v} = -\frac{\Ri_b\Rey^2\Sc\beta^2}{\varepsilon}
    \mathrm{e}^{-(1+y)/\varepsilon}\widetilde{\phi},\label{eq:high vs odes 1}\\
    \left(
    L^2 - \frac{1}{\varepsilon}\frac{\mathrm{d}~}{\mathrm{d}y}
    \right)\widetilde{\phi} = -\frac{2}{\varepsilon}\widetilde{v},\label{eq:high
    vs odes 2}%
\end{gather}
\end{subequations}
with $\mathrm{d}\widetilde{\phi}/\mathrm{d}y=0$ for $y=\pm 1$
replacing the no-flux condition at the walls.

Away from the boundary, where $1 + y \gg \epsilon$, the velocity mode decouples
from the sediment, since Eq.~\eqref{eq:high vs odes 1} reduces to
$L^4 \widetilde{v} = 0$. 
This is integrated with respect to the two upper wall boundary conditions on
$\widetilde{v}$, plus $\widetilde{v}(y)=0$ at $y=-1$, which is required for
the solution to match with the behaviour at the lower wall.
The resulting expression may be written as
\begin{equation}
    \widetilde{v}(y) = B\left\{
        \left(\frac{y}{2\beta} + b\right)
        \sinh[\beta(y-1)]
        - \left( \frac{1}{2} + b\beta \right)
        (y-1)\cosh[\beta(y-1)]
        \right\},
    \label{eq:outer soln}%
\end{equation}
where $b = (\tanh 2\beta+ 2\beta)/[2\beta(\tanh 2\beta - 2\beta)]$ and $B$ is an
undetermined constant due to the linearity of the problem.
Expanding Eq.~\eqref{eq:outer soln} at $y=-1$ shows that this solution
scales like $\widetilde{v}\sim y$ as it approaches the lower wall. 

Near the boundary, we use the rescaled vertical coordinate from
\S\ref{sec:high vs}, defining $Y = (y + 1) / \varepsilon$.
The, we recast Eqs.~\eqref{eq:high vs odes 1} and~\eqref{eq:high vs odes 2} in
terms of the variables $\widetilde{V}(Y) = \widetilde{v}(y)/\varepsilon$ and
$\widetilde{\Phi}(Y) = \widetilde{\phi}(y)$, giving
\begin{subequations}
\begin{gather}
    \frac{\mathrm{d}^4 \widetilde{V}}{\mathrm{d}Y^4}
    = \varepsilon^2\beta^2\left(\frac{\mathrm{d}^2\widetilde{V}}{\mathrm{d}Y^2}-\Ri_b \Rey^2
    \Sc\,\mathrm{e}^{-Y}\widetilde{\Phi}
    -\varepsilon^2\beta^2\widetilde{V}\right),\label{eq:inner odes 1}\\
    \frac{\mathrm{d}^2\widetilde{\Phi}}{\mathrm{d}Y^2} -
    \frac{\mathrm{d}\widetilde{\Phi}}{\mathrm{d}Y} = 
    \varepsilon^2(\beta^2 \widetilde{\Phi} - 2\widetilde{V}).
    \label{eq:inner odes 2}%
\end{gather}
\end{subequations}
Matching with the outer Eq.~\eqref{eq:outer soln} solution, as well as
Eq.~\eqref{eq:high vs odes 2}, which implies that
$\mathrm{d}\widetilde{\phi}/\mathrm{d}y = 2 \widetilde{v}$ away from the wall,
requires that $\widetilde{V}\sim Y$ and
$\mathrm{d}\widetilde{\Phi}/\mathrm{d}Y\sim 2\varepsilon^2 \widetilde{V}$ as
$Y\to\infty$. Therefore, the right-hand side of Eq.~\eqref{eq:inner odes 2} is
$O(\varepsilon^2)$ and the only possible lower-order contribution to
$\widetilde{\Phi}$ is a constant. 
The solution for $\widetilde{V}$ depends upon whether this constant is nonzero.

First, we assume it is nonzero and write $\widetilde{\Phi}(Y) =
\widetilde{\Phi}_0 + \varepsilon^2\widetilde{\Phi}_2(Y) + \ldots$.
When this is substituted into Eq.~\eqref{eq:inner odes 1}, it becomes
clear that for $\widetilde{V}$ to possess an $O(1)$ component (as required for
the matching condition), $\Ri_b\Rey^2\Sc\beta^2 = O(\varepsilon^{-2})$.
In this case, to leading order, Eqs.~\eqref{eq:inner odes 1} and~\eqref{eq:inner
odes 2} can be written as
\begin{subequations}
\begin{gather}
    \frac{\mathrm{d}^4 \widetilde{V}}{\mathrm{d}Y^4} = -\Omega
    \widetilde{\Phi}_0 \mathrm{e}^{-Y},\label{eq:inner reduced 1}\\
    \frac{\mathrm{d}^2\widetilde{\Phi}_2}{\mathrm{d}Y^2} -
    \frac{\mathrm{d}\widetilde{\Phi}_2}{\mathrm{d}Y} = 
    \beta^2\widetilde{\Phi}_0 - 2\widetilde{V},\label{eq:inner reduced 2}
\end{gather}
\end{subequations}
where $\Omega = \varepsilon^2\beta^2\Ri_b\Rey^2\Sc = O(1)$.
Integrating Eq.~\eqref{eq:inner reduced 1} with respect to $\widetilde{V}(0) =
\mathrm{d}\widetilde{V}(0)/\mathrm{d}Y = 0$ and enforcing the matching condition
for $\tilde{V}$ leads to the solution
\begin{equation}
    \widetilde{V}(Y) = \Omega \widetilde{\Phi}_0 (1 - Y - \mathrm{e}^{-Y}).
    \label{eq:high vs v mode}%
\end{equation}
We can then apply a solvability criterion on Eq.~\eqref{eq:inner reduced 2} to
constrain $\Omega$ further. Since $\mathrm{e}^Y$ lies in the kernel of the
operator $\mathrm{d}^2/\mathrm{d}Y^2 - \mathrm{d}/\mathrm{d}Y$, the Fredholm
alternative implies that
\begin{equation}
    \int_0^{\infty} \mathrm{e}^{-Y}(\beta^2 \widetilde{\Phi}_0 - 2\widetilde{V})
    \,\mathrm{d}Y = 0,
\end{equation}
for Eq.~\eqref{eq:inner reduced 2} to have a solution. Evaluating
the integral implies that $\Omega = -\beta^2$.
Therefore, we conclude that the marginal stability boundary is given by
\begin{equation}
    \Ri_b = -\frac{1}{\Rey^2 \Sc}\left(\frac{v_s}{\kappa}\right)^{\!2}
    = -\Sc\, v_s^2.
\end{equation}
The solution in Eq.~\eqref{eq:high vs v mode} is unique, up to
$\widetilde{\Phi}_0$, which must be selected in concert with $B$ to complete the
matching procedure. This determines the primary mode for each $\beta$.  Note
that the corresponding stability boundary is independent of $\beta$ and, in the
high $v_s/\kappa$ limit, agrees
favourably with the mode computations in the full system, as
plotted in Fig.~\ref{fig:lin stab}.

Alternatively, if $\widetilde{\Phi}_0 = 0$, Eq.~\eqref{eq:inner odes 1}
requires that $\Ri_b\Rey^2\Sc\beta^2 = O(\varepsilon^{-4})$ for $\widetilde{V}$ to
have an $O(1)$ component. We repeat the previous procedure, defining $\Gamma =
\varepsilon^4\Ri_b\Rey^2\Sc\beta^2 = O(1)$, then writing out
Eqs.~\eqref{eq:inner odes 1} and~\eqref{eq:inner odes 2} to leading order.
In this case, we combine them to give
\begin{equation}
    \frac{\mathrm{d}^6\widetilde{\Phi}_2}{\mathrm{d}Y^6}
    - \frac{\mathrm{d}^5\widetilde{\Phi}_2}{\mathrm{d}Y^5}
    = 2\Gamma \widetilde{\Phi}_2\mathrm{e}^{-Y}\!,
    \label{eq:Phi2 ode}%
\end{equation}
subject to the boundary conditions $\mathrm{d}\widetilde{\Phi}_2/\mathrm{d}Y =
\mathrm{d}^2\widetilde{\Phi}_2/\mathrm{d}Y^2 =
\mathrm{d}^3\widetilde{\Phi}_2/\mathrm{d}Y^3 = 0$ at $Y = 0$ (where the final two
constraints are derived from the bottom wall conditions on $\widetilde{V}$).
Additionally, the matching condition requires that
$\mathrm{d}\widetilde{\Phi}_2/\mathrm{d}Y \sim Y$ as $Y\to\infty$.
Equation~\eqref{eq:Phi2 ode} is a differential eigenvalue problem.
The eigenvalues, ordered according to ascending magnitude, correspond to the secondary and
higher modes.
We index them by $\Gamma = \Gamma_j$ for $j = 2, 3, \ldots$.
For each mode, the marginal stability is therefore given by
\begin{equation}
    \Ri_b = \frac{\Gamma_j}{\Rey^2 \Sc \beta^2}\left(
    \frac{v_s}{\kappa}\right)^{\!4}\!.
\end{equation}
We solve Eq.~\eqref{eq:Phi2 ode} numerically to compute $\Gamma_2 = -0.656$ for
the secondary modes and include the corresponding curves on Fig.~\ref{fig:lin
stab} (for the two $\beta$ of lowest frequency), which ultimately agree with the
data for the full problem when $v_s/\kappa \gg 1$.

\end{document}